\documentclass[preprint,floats,aps,showpacs,epsf]{revtex4}
\usepackage{amssymb}
\usepackage{latexsym}
\usepackage{amsmath}
\usepackage{epsfig}
\usepackage{color}
\usepackage{graphicx}
\usepackage{hyperref}
\numberwithin{equation}{section}
\numberwithin{figure}{section}
\numberwithin{table}{section}
\newcommand{\be}{\begin{equation}}
\newcommand{\ee}{\end{equation}}
\newcommand{\bea}{\begin{eqnarray}}
\newcommand{\eea}{\end{eqnarray}}

\newcommand{\ri}{i}
\newcommand{\re}{\mbox{e}}

\begin{document}
\title{Bound-States Dynamics in One-Dimensional Multi-Species Fermionic Systems }

\author{ P. Azaria$^{1,2}$}
\affiliation{ $^{1}$  Laboratoire de Physique Th\'eorique de la Mati\`ere Condens\'ee, Sorbonne Universit\'es, UPMC Univ Paris 06 and CNRS,  4 Place Jussieu, 75005 Paris, France\\
$^{2}$ Physics Department, Technion, 32000 Haifa, Israel}
\date{\today}

\begin{abstract}
 In this work we  provide for a description of the low-energy physics of interacting multi-species fermions in terms of the bound-states that are stabilized in these systems when  a spin gap opens. We argue that,  at energies much smaller than the spin gap,  these systems are described by a Luttinger liquid of bound-states that depends, on top of the charge stiffness  $\nu$ and the charge  velocity $u$, on a "Fermi"
momentum $P_F$ satisfying $qP_F = Nk_F$ where $q$ is the charge of the bound-state, $N$ the number of species and $k_F$
is the Fermi momentum in the non-interacting limit. We further argue that for generic interactions, generic bound-states are likely to be stabilized. They are  associated with emergent, in general non-local,  symmetries and are in the number of five. The first two  consist  of   either a charge $q=N$ local $SU(N)$ singlet or a charge $q=N$ bound-state made of two local $SU(p)$ and $SU(N-p)$ singlets. In this case the Fermi momentum $P_F=k_F$ is preserved. The three others have an enhanced Fermi vector $P_F$. The latter are either charge $q=2$ bosonic p-wave and s-wave pairs with $SO(N)$ and $SP(N)$ symmetry and $P_F=Nk_F/2$ or a composite fermion of charge $q=1$ with  $P_F=Nk_F$. The instabilities of these Luttinger liquid states towards incompressible phases and their possible topological nature  are also discussed. 

\end{abstract}

\maketitle

%
%
\section{ Introduction}  
As is well known the Luttinger liquid constitutes the universality class of a large number of gapless quantum systems in one dimension\cite{haldane, luttinger}. Spinless bosons or fermions on a lattice, spin models like the XXZ spin chain\cite{giamarchi, pham}, edge states in the FQHE\cite{wen, fisherglazman} are all well known examples of one dimensional  systems which are described by the Luttinger liquid theory. The Luttinger liquid  is  also expected to describe the low-energy sector of more involved models with $N$  species  of particles: when a gap is present  in the species, or spin, sector the low-energy physics is expected to be captured by the total charge, or density, fluctuations which are described by the Luttinger liquid hamiltonian. Examples can be found, among others, in electronic ladders\cite{gogolin, balents, fabrizio, orignacgiamarchi} or  cold-atoms systems\cite{ho, phle, capponi} with hyperfine spin $F=(N-1)/2$.  All these systems have in common that their low energy physics  depend on two Luttinger  parameters, a stiffness $\nu$ and a velocity $u$. These parameters, which may be eventually taken as phenomenological input parameters, completely determine the asymptotics of the correlation functions of physical observables.   Does this mean that the low energy physics of these systems is the same?  As we shall argue in this work, though this is certainly  true as far as particle-hole (or plasmon) excitations are involved, the  nature of the elementary excitations in these Luttinger liquids is  different. 

Indeed  when a gap opens in the spin sector single particle correlation functions fall off exponentially and only certain singlet combinations  remain massless. These combinations are bound-states of the elementary fermions  and are, at energies $E$ much smaller than the spin gap $\Delta$, the relevant excitations of these systems.   A celebrated  example is that of the stabilization of bosonic BCS pairs by the opening of a spin gap in the $S=1/2$ attractive Hubbard model\cite{giamarchi}. Many other examples with more than two components were also reported in the literature. In cold atom problems, s-wave pairs made of  hyperfine spins $F>1/2$ singlets, as well as trionic or quartet bound-states made of $SU(N)$ $(N=3,4)$ singlets, were also shown to exist\cite{phle, Wu, capponi2, capponi, roux2008, azaria2, rapp, dukelsky2008, ulbricht}. 

The purpose of this work will be to present a  description 
of the dynamics of these bound-states within the framework of the Luttinger liquid theory and to provide for a common view  of the low energy physics of multi-species fermionic systems when a spin gap is present. As we shall see, the  Luttinger liquid theory offers a natural framework to describe bound-states. Indeed, Luttinger liquids may be distinguished by their non-zero  charge $Q$ and current $J$ spectrum or zero-mode spectrum. When a bound-state of charge $q$ is stabilized by the spin gap,   the total charge $Q$ is to be quantified in units of the elementary bound-state charge $q \in \mathbb{N}$ with $Q=n q$. The fundamental excitation of charge $q$, which is either a boson when  $q$  is even or a fermion when  $q$ is odd, is the minimal charge that one can add (remove) to (from) the system and play an analogous role as the electron in a one species system. Similar considerations yield to the quantization of the current $J=m j$ for fermions and $J=2 m j$ for bosons where $j \in \mathbb{N}$ and  $2 j \in \mathbb{N}$ are the minimal  non-zero currents both   systems can support. Therefore, in order to compleetly characterize the bound-state Luttinger liquid state, one needs, on top of the Luttinger parameters $u$ and $\nu$, to specify the elementary charge and current quantum numbers $(q,j)$. A bound-state Luttinger liquid can  then be viewed as an additional selection rule on the zero mode spectrum  ($Q$, $J$). The latter selection rules  keep track of the underlying possible orders in the high-energy spin sector.  

 As a first result,   we shall see in the section (II) that,  independently of the nature of the high-energy physics involved,  the bound-state quantum numbers $(q,j)$ are not arbitrary. For instance, with the additional assumption that the bound-states are {\it local} in terms of the elementary fermions, we find that  they have to be dual in the sense $qj=N$. Hence, a bound-state Luttinger liquid is characterized, on top of the Luttinger parameters $\nu$ and $u$, by the charge $q$ and the  current $j$ quanta solutions of the latter constraint. Owing to  the relation between current and momentum one can associate a momentum scale to the bound-states, $P_F=j k_F$, where $k_F$ is the Fermi momentum of the elementary fermions, and consequently rewrite the constraint as $qP_F=Nk_F$. 
Of course, at some point, the specific nature of the ordering in the spin sector should come into play and select specific values of $q$ and $j$. We shall see in the section (III), that under the assumption of dynamical symmetry enlargement in the spin sector, some generic bound-states, i.e. particular values of $(q,j)$, are likely to be stabilized for generic hamiltonians. The latter  are in the number of five and are associated with emergent duality symmetries. The first two  types of bound-states have  $(q, j) = (N, 1)$ and are either a $SU(N)$ singlet or a bound-state made of two $SU(p)$ and $SU(N-p)$ singlets with $1\le p<N$. The three other types are, either  $SP(N)$ singlet s-wave and   $SO(N)$  singlet p-wave bosonic states with  $(q, j) = (2, N/2)$ or   $SO(N)$ singlet  composite fermions with $(q, j) = (1, N)$. In the section (IV) we shall give explicit forms of the associated wave functions and give their expressions in terms of the elementary fermionic species. We shall show, using a low-energy approach, that after averaging over the gapped spin degrees of freedom,  they  have a finite overlap with the single particle  creation operator of the Luttinger liquid.
After having characterized these generic bound-states, we shall finally investigate the instabilities of the corresponding Luttinger liquid states toward possible incompressible phases
in the section (V).  As one of the consequences of the bound-state dynamics we shall find that,  since it is $P_F=j k_F $ and not $k_F$ that controls the commensurability effects with the lattice, when $j>1$  possible non-degenerate Mott phases with topological order might be stabilized for systems. We finally conclude in section (VI) where we  discuss open problems and further directions of works.

In the following we shall consider systems with $N$ species of fermions  on a one-dimensional lattice of length $L$ with a generic hamiltonian:
\be
{ H} = -t \sum_{j, a} [c^{\dagger}_{a,j} c^{}_{a, j+1} + c^{\dagger}_{a, j+1} c^{}_{a, j}]
+  V_{\rm int}(c^{\dagger}_{a,i}, c^{}_{b,j})
\label{hamiltoniangeneric}
\ee
where the operators $c^{\dagger}_{a,j}$ create a fermion of species $a=(1,...,N)$ at lattice site $j$ and  are subject to periodic boundary conditions: $c^{\dagger}_{a,j+L}=c^{\dagger}_{a,j}$. 
We assume that the potential $V_{\rm int}$ is short range, translationally and parity invariant,  and preserves  the total number of particles.  We  shall  further assume that a  gap $\Delta$ opens  everywhere in the spin sector and that  the system remains massless.  The interaction pattern between the species is supposed to be in such  a way  that none of them decouple;  if the system decouples into two or more subsets, then we consider applying  the analysis to each one of these separately.  For simplicity  a balanced incommensurate density per species ${\bar \rho}_a = {\bar \rho}= {\cal N}/L$ is also assumed so that  there is only one Fermi momentum $k_F=\bar \rho \pi$.
%
%

\section{ Bound-state Luttinger liquids }

Our approach is a low-energy one in which the electron operators
$c_{a,j}$, $a=(1,...,N)$,  decompose into  left and right components as:
\be
c_{a,j}/\sqrt{a_0} \sim \re^{-ik_F x} \psi_{a,L}(x) + \re^{ik_F x} \psi_{a,R}(x).
\label{chiralfermions}
\ee
where $x=ja_0$  ($a_0$ being the lattice spacing) and $k_F= \pi \bar \rho$ is the Fermi momentum associated with each species.
The above right and left fermions can be in turn expressed in terms of two dual bosonic fields\cite{gogolin} $\theta_a$ and $\phi_a$ satisfying: $[\phi_a(x), \theta_b(y)] = i \delta_{ab} Y(y-x)$, where $Y(u)$ is the step function ($Y(0)=1/2$). We have
\be
\psi_{a,L(R)} =  \frac{\kappa_a}{\sqrt{2 \pi }} \;  \re^{ \displaystyle{-i\sqrt{\pi}[\theta_a \pm \phi_a]}},
\label{bosonization}
\ee
where the $\kappa_{a=1,...,N}$ are anticommuting Klein factors, $\{\kappa_a, \kappa_b\}= 2\delta_{ab}$, that insure the anticommutation between fermions of different species. For each species, the bosonic fields  $\theta_a$ and $\phi_a$ are related to the current densities, $j_a(x)=\partial_x \theta_a/\sqrt{\pi}$,  and uniform particle densities (relative to the ground state)  $\rho_a(x) =  \partial_x \phi_a/\sqrt{\pi}$. 
The zero modes of the charge and current densities, ${ Q}_a=\int dx \; \partial_x \phi_a/\sqrt{\pi}$ and  ${ J}_a= \int dx\;  \partial_x \theta_a/\sqrt{\pi}$,  associated with each species
\bea
Q_a &=& \int dx [\psi_{a,L}^{\dagger} \psi_{a,L}^{}+ \psi_{a,R}^{\dagger} \psi_{a,R}^{}] ,  \nonumber \\
J_a &=& \int dx [\psi_{a,L}^{\dagger}\psi_{a,L}^{} - \psi_{a,R}^{\dagger}  \psi_{a,R}^{}], 
\label{QaJa}
\eea
are topological quantities which, as  befits from charge quantization,   are integers. In a system with periodic boundary conditions they  are subjected to the additional constraint\cite{pham, giamarchi}
\be
 (Q_a \pm J_a)\in 2\mathbb{Z} \; {\rm even}, \; a=(1,...,N). 
\label{periodic}
\ee
When a gap $\Delta$ is present in spin  space, we expect the low-energy physics  to be captured  by a Luttinger liquid describing the fluctuations of the {\it total} charge and current densities of the system described by the bosonic fields
\be
\Phi_c = \frac{1}{\sqrt{N}} \sum_{a=1}^{N}\phi_a, \;
\Theta_c = \frac{1}{\sqrt{N}} \sum_{a=1}^{N} \theta_a,
\label{PhiTheta}
\ee
where  $[\Phi_c(x), \Theta_c(y)] = i  Y(y-x)$.
Integrating out the spin degrees of freedom, the effective hamiltonian at scales $E << \Delta$ is therefore expected to be\cite{haldane}
\be
{\cal H}_{} = \frac{u}{2} \int dx \; [\frac{1}{K} (\partial_x \Phi_c)^2 +
K (\partial_x \Theta_c)^2],
\label{luttinger}
\ee
where $u$ is a velocity and $K$ is the  Luttinger parameter that measures  the interaction between the elementary fermions. Seemingly,    the spin degrees of freedom only affect the parameters $u$ and  $K$ which anyhow depend in a non-universal way on the details of the microscopic hamiltonian and can be taken as phenomenological input parameters. The underlying spin order though,  have a nontrivial effect on the topological excitations associated with  the zero mode part of the bosonic fields $\Phi_c$ and $\Theta_c$
\be
Q =\sum_{a=1}^N Q_a, \; J = \sum_{a=1}^N J_a,
\label{zeromode}
\ee
or $Q= \sqrt{\frac{N}{\pi}}  \int dx \; \partial_x \Phi_c$, $
 J =  \sqrt{\frac{N}{\pi}} \int dx \; \partial_x \Theta_c$.
Indeed, as discussed above, when a spectral gap $\Delta$ opens  in the spin sector, excitations  involving arbitrary non-zero values of the charge and current operators (\ref{QaJa}), ${ Q}_a$ and ${ J}_a $,   are also  gapped in general. Only certain  singlet combinations of the elementary fermions   survive at low energies and remain massless. This  restricts, on top  of the constraints (\ref{periodic}), the allowed eigenvalues of both $Q$ and $J$ zero mode operators  (\ref{zeromode}). Taking into account   the constraint (\ref{periodic}), we find suitable to parametrize  the bound-states  with help of two integers $(q,j)$ as
\bea
&q& \;  {\rm even } \hspace{1.cm}  Q=n \; q, \; 
 J=2 m \; j,  
\label{bosoqm}
\\
 &q& \;  {\rm odd } \hspace{1.1cm}  Q=n \; q, \; J= m\;  j, 
\label{fermiqm} \\
& & \, \hspace{1.9cm} (n q \pm m j) \; {\rm even},
\nonumber 
 \eea
where $(n,m)$ are relative integers. In the latter equation, 
(\ref{bosoqm}) and (\ref{fermiqm}) are bosonic and fermionic
solutions respectively. The two quantities $(q, j)$ are in fact not independent. Let us consider indeed the vertex operator that creates a state with charge $Q$ and  current $J$
\be
V_{Q, J} \equiv \exp{  \left[\displaystyle{i\sqrt{\frac{\pi}{N}}[ Q \Theta_c + J \Phi_c]}\right]},
\label{vertex}
\ee
and look at  the (imaginary time)  correlation function, 
\be
\langle V_{Q_1 J_1}(x_1, \tau_1) V_{Q_2 J_2}(x_2, \tau_2) \rangle =
 |z_1 - z_2|^{\Delta_{12}} \re^{-i \Theta_{12} \Gamma_{12}},
\ee
where  $z=\tau + i x/u$  and $\Theta_{12}= {\rm Arg}(z_1-z_2)$. We find for  $q$ odd  $\Gamma_{12} = q j (n_1m_2 + m_1 n_2)/2N$, $\Delta_{12} = (n_1 n_2 q^2 + m_1 m_2 K^2 j^2)/2KN$ while for $q$ even $\Gamma_{12} = q j (n_1m_2 + m_1 n_2)/N$, $\Delta_{12} = (n_1 n_2 q^2 +4 m_1 m_2 K^2 j^2)/2KN$.
 Analicity of the correlation function in the complex plane\cite{Tsvelik} requires 
 $\Gamma_{12}$ to be an integer which, using the constraints
(\ref{fermiqm}) and  (\ref{bosoqm}), implies that
$
q j = l N
$
where $l$ is an arbitrary integer.  As we shall see below, only the case with $l=1$  corresponds to {\it local} bound-states when expressed in terms of the elementary fermions.  We believe that these are the  states that can be stabilized with a hamiltonian of the kind (\ref{hamiltoniangeneric}) and from now on  we shall focus on the sets of the bound-state  solutions $(q, j)$ of
\be
 q j = N.
\label{bssolution}
\ee
We shall comment briefly later on the $l\neq1$ states.
To get some physical insight of the meaning of
(\ref{bssolution}) we notice that in a Luttinger liquid, the  quantum of current $j$ defines a momentum scale $P_F=j k_F$  which, due to (\ref{bssolution}), must satisfy
\be
q P_F =  N k_F.
\label{luttheo}
\ee
For fermions  we may interpret (\ref{luttheo}) as an   extented   Luttinger theorem\cite{luttingertheorem, yamanaka} in one dimension  when a spin gap is present. 
 For bosonic bound-states $2P_F$ (for practical purpose we use the same symbol for fermion and bosons) governs the period of the oscillations of the charge density wave and is related to  the bound-state density by $P_F = \pi \rho_{{\rm BS}}$. The constraint (\ref{luttheo}) then yields
for the bound-state density
\be
 \rho_{{\rm BS}} =  N \bar \rho/q, 
\label{bsdensity}
\ee 
which is the one that we  would calculate in  a limit where  the bosonic bound-state are free hard-core bosons.

The constraint (\ref{bssolution}) (or equivalently (\ref{luttheo})) is not trivial.  For instance, the solutions with $j> 1$ are reminiscent
of some degree of confinement of the current 
which, as we shall see in the next section, is the signal that the non-zero momenta components
of the bound-states might involve composites of particle-hole of the elementary fermions.  For the time being let us comment qualitatively on the solutions of (\ref{bssolution}).  Given  the number $N$ of species,  the possible bound-state solutions ($q$, $j$)  are strongly constrained  by  $j=N/q\in \mathbb{N}$. First we find that only bosonic bound-states exist for even $N$ while  for odd $N$ they are fermions. For  example,  we find that  in the simplest case of $N=2$ the only bound-state solution is given by $(q=2, j=1)$ which corresponds to  spin $F=1/2$ BCS pairs. In the case of   $N=3$ we get two solutions: $(q=3, j=1)$ and $(q=1, j=3)$. The first solution corresponds to a charge $q=3$ fermionic trionic bound-state\cite{azaria2, rapp} with a preserved Fermi momentum  at $P_F=k_F$. 
The second bound-state solution with charge $q=1$  displays an enhanced Fermi surface, $P_F=3k_F$. As we shall see, this bound-state is   a composite fermion  made of two particles and one hole. For $N=4$ there are two solutions with $(q=4, j=1)$ and $(q=2, j=2)$. These are  bosonic quartet bound-states and  spin $F=3/2$ BCS pairs\cite{roux2008}. More solutions can be found for higher values of $N$ with or without an enhancement of the Fermi momentum. 

\subsection{Universal Description of the Bound-State Luttinger Liquids}
Though in general the different bound-state solutions describe different physics, they can be  described by the same effective bosonic theory provided one uses suitable rescaled fields. Introducing  new bosonic fields $\bar \phi$ and $\bar \theta$ with help of the canonical transformation
\be
\bar \phi= \frac{\sqrt{N}}{q} \; \Phi_c, \; 
\bar \theta = \frac{\sqrt{N}}{j} \; \Theta_c,
\label{barphitheta}
\ee
the Luttinger liquid hamiltonian (\ref{luttinger}) can be brought into the universal form:
\be
{\cal H} = \frac{u}{2} \int dx \; [\frac{1}{\nu} (\partial_x \bar \phi)^2 +
\nu (\partial_x \bar \theta)^2], 
\label{luttingerbar}
\ee
where
\be
\nu = NK/q^2.
\label{luttparabs}
\ee
The latter relation shows that if $K$ is a measure of the interaction between the  species, $\nu$ measures the interaction between the bound-states. In particular $\nu$ is the Luttinger parameter that controls the power-law behaviors of the different correlation functions of the system.

The charge and current operators (\ref{zeromode}), $Q$ and $J$, express in terms of  the  "dimensionless" (i.e. independent of both $q$ and $j$) charge and current
\be
Q = q\;  \bar Q \; , \, J= j\;  \bar J,
\label{qjbar}
\ee
where $
\bar Q = \int \partial_x \bar \phi(x)/\sqrt{\pi} \; {\rm and} \; \bar J = \int \partial_x \bar \theta(x)/\sqrt{\pi} 
$.  
With use of (\ref{barphitheta}) and   depending on the parity of $q$  the conditions (\ref{bosoqm},\ref{fermiqm}) are now
\bea
&q& \;  {\rm even } \hspace{1.cm}  \bar Q=n  \; 
 \bar J=2 m,   
\label{bosobar}
\\
 &q& \;  {\rm odd } \hspace{1.1cm}  \bar Q=n, \; \bar J= m, \;  
\label{fermibar} \\
& & \, \hspace{1.9cm} (n  \pm m ) \; {\rm even}.
\nonumber 
 \eea
The latter constraints are the ones defining both bosonic and fermionic Luttinger liquids\cite{pham} and 
the hamiltonian (\ref{luttingerbar}) describes the low energy physics of spinless bosons or fermions with periodic boundary conditions. Once a bound-state solution (\ref{bssolution}) is given in terms of $(q, j)$ the low energy dynamics of the bound-state Luttinger liquid is that of a charge $q$  boson with density $\rho_{{\rm BS}}$ (\ref{bsdensity}) or of a charge $q$ spinless fermion with Fermi momentum 
\be
P_F = \pi  \rho_{{\rm BS}}.
\label{fermivector}
\ee
The relations (\ref{barphitheta}) and (\ref{qjbar}) allow to translate all known results for the spinless fermionic and bosonic Luttinger liquid\cite{giamarchi, pham}. In particular,  the relevant physical operators can be expressed in the basis of the vertex operators
\be
V_{\bar Q, \bar J}\equiv \exp{i\sqrt{\pi}[\bar Q \bar \theta + \bar J \bar \phi]},
\label{vertexbar}
\ee 
which carry physical charge $q\;  \bar Q$, current $j\;  \bar J$
and momentum $ \bar J P_F$. For instance both bosonic and fermionic bound-states single particle creation operators  have charge $\bar Q=1$ and are  given   by\cite{haldane2, giamarchi}
\be
\Psi^{\dagger}_{B/F} \simeq \sum_{\bar J {\rm even}/ {\rm odd} } \alpha_{\bar J}\;  \re^{i [\bar J P_F x +\bar J \sqrt{\pi}  \bar \phi +\sqrt{\pi}  \bar \theta]}.
\label{bosonfermionexpansion}
\ee
Similarly the density operator (relative to the ground state) is given 
by
\be
\rho(x) \simeq  \frac{q}{\sqrt{\pi}} \partial_x \bar \phi  +
 \sum_{\bar J {\rm even}} \beta_{\bar J } \; \re^{i \bar J [P_F x + \sqrt{\pi} \bar \phi ]}.
\label{densitybsboson}
\ee
In both the latter espressions, the constants $\alpha_{\bar J}$ and $\beta_{\bar J}$ are non-universal  and depend on the details of the microscopic hamiltonian. To leading order the boson operator is given by its $P=0$  component
\be
\Psi^{\dagger}_{B} \simeq \exp{[i \sqrt{\pi} \bar \theta]},
\label{leadingboson}
\ee
while the fermionic bound-state creation operator has leading components at $P=\pm P_F$
\be
\Psi^{\dagger}_{F} \simeq  \Psi^{\dagger}_{ L}\re^{iP_Fx} + \Psi^{\dagger}_{ R} \re^{-iP_Fx},
\label{leadingfermionexpansion}
\ee
where
\be
\Psi^{\dagger}_{F, L(R)} \simeq \exp  \left[\displaystyle{i\sqrt{\pi}[   \bar \theta \pm  \bar \phi]}\right].
\label{minimalfermion}
\ee
In both the bosonic and fermionic cases the phase diagram
of the Luttinger liquid is well known\cite{giamarchi} and depends on $\nu$. From the long distance behavior of the equal-time correlation functions $<\Psi^{\dagger}_{B}(x) \Psi^{}_{B}(0)> \sim x^{-1/2\nu}$;  $<\Psi^{\dagger}_{F}(x) \Psi^{}_{F}(0)> \sim
x^{-(\nu+\nu^{-1})/2} \cos{(P_Fx)}$ and $<\rho(x) \rho(0)> \sim
x^{-2\nu} \cos{(2P_Fx)}$ one deduces that the dominant instability
is given by the bound-state/bound-state correlation function
for $\nu >1/2$ and $\nu >1/\sqrt{3}$ in both bosonic and fermionic cases. For stronger repulsions and smaller values of $\nu$ the dominant instability becomes eventually of the charge density wave type at the  wave vector $2P_F$. We notice though, that even in this regime of couplings, as far as the energy scale is much smaller than the spin gap, the bound-state description is still sensible. In particular,   the bound-state manifest themselves in the nontrivial wave vector $2P_F$.
At this point let us stress that due to  (\ref{luttparabs}),  a repulsive interaction between the bound-states  ($\nu <1$) might results from either  attractive or repulsive interactions between the  elementary fermionic specy  ($K>1$ or $K<1$). In particular
both (\ref{leadingboson}) and (\ref{minimalfermion})
are the operators for a free hard-core boson or a fermion in the limit $\nu=1$ which does not corresponds in general to $K=1$ 
(except when $q=j$).

Let us close this section by commenting on the solutions
with $l>1$. The constraint on the Fermi vector 
is then given by $q P_F= l N k_F$ instead  of (\ref{luttheo}). In this case, as we shall argue, the bound-states are non local objects when expressed in terms of the elementary fermions.  To see this let us  consider as an example fermionic bound-states solutions ($l N$ odd). Then one may bring the Luttinger hamiltonian to its universal form (\ref{luttingerbar}) with, insead of (\ref{barphitheta}), the canonical transformation
$
\bar \phi= \sqrt{N}  \Phi_c/q, 
\bar \theta = l \sqrt{N} \Theta_c/j.
$
As a consequence the values of the current $\bar J$ are {\it still} constrained  to be $\bar J = l m$, $m \in \mathbb{Z}$. In particular the dimensionless vertex operator (\ref{vertexbar}) that creates a state with charge $q$ and current $\pm j$ is given by
\be
V{}_{1, \pm l}=\exp{i\sqrt{\pi}[ \bar \theta \mp  l \bar \phi]}.
\label{vertexbarl}
\ee 
This state is a composite fermion and is non-local when expressed
in terms of the original fermions. In the simplest case of a single species $N=1$ with charge $q=1$ and current quantum number $j=l$ we have: $V{}_{1,  -l}(x) \sim \psi^{\dagger}_R (\psi^{\dagger}_R\psi^{}_L)^{(l-1)/2}$ and $V{}_{1,  l}(x) \sim \psi^{\dagger}_L (\psi^{\dagger}_L\psi^{}_R)^{(l-1)/2}$.
As is well known  when  $\nu =1/l$ these states identify
with the electron operator at the edges of FQH device\cite{wen, stonefisher, kane}. As we shall see, when $l=1$,  a the bound-state
with $(q=1, j=N)$ may be still a composite fermion but can be  made local thanks to the $N$ spin degrees of freedom.
%
%
\subsection{ Stiffnesses, Luttinger Parameters and Transport Properties}
Thanks to the relation (\ref{bssolution}) a bound-state Luttinger
liquid is characterized by the three quantities $(P_F, u, \nu)$ or equivalently ($q, u, \nu)$. These independent parameters  could be in principle extracted  from the knowledge of the different stinffnesses of the problem\cite{cazalilla}. 
As is well known the  two Luttinger  parameters $u$ and $\nu$  can be related to different stiffnesses or rigidities associated with ground state properties of the system\cite{cazalilla}. The first stiffness is related to the macroscopic compressibility at zero temperature which can be related to  the second derivative of the ground state energy ${\cal E}_0$ with respect to the  total number of particles:
\be
\kappa^{-1} = L \frac{\partial^2 {\cal E}_0}{\partial {\cal N}^2} = \nu q^2/\pi u.
\label{inversecompressibility}
\ee
The other stiffness is the zero temperature phase stiffness $D_{\alpha}$ which is related to the response of the system to an infinitesimal twist $\alpha$ in the boundary conditions: $c^{\dagger}_{a,i+L}  = \re^{i\alpha} c^{\dagger}_{a,i}$. In a Luttinger Liquid the ground state energy in the presence of the twist $\alpha$  is to be found in  the reduced space with total zero-charge  (and zero particle-hole excitations) described by  the projected hamiltonian 
\be
 {\cal H}(\alpha) = {E}_0 + \frac{1}{2L} D_{\alpha} ( \frac{\pi}{q} \bar J - \alpha)^2,
\ee
where $\bar J$ is the dimensionless current operator (\ref{qjbar})
and 
\be
D_{\alpha}= L \frac{\partial^2 {\cal E}_0}{\partial \alpha^2} = u \nu q^2/\pi.
\label{twiststiffeness}
\ee
is the phase stiffness associated with the capability of the system to sustain a persistent current. From (\ref{inversecompressibility})
and (\ref{twiststiffeness}) we see that $u$ and $\nu$ can be obtained from $\kappa$ and $D_{\alpha}$ only if  {\it  $q$ is known}. To determine the value of the bound-state charge $q$ one has to consider the full dependence of the ground state energy on the twist $\alpha$\cite{yang, loss}. As  $\bar Q=0$, and wathever the parity of $q$ is,  $\bar J$  has to be even. 
Therefore the ground state energy ${\cal E}_0(\alpha)$  is a periodic function  with period $2\pi /q$ and has minima at $\alpha_m= 2\pi m/q$, $m\in \mathbb{Z}$. The corresponding eigenstates have quantum numbers $\bar J =2m$ and carry   persistent currents $J=2m j$. We thus find that by varying $\alpha$ in the interval $[0,2\pi[$ the ground state energy ${\cal E}_0(\alpha)$ has exactly $q$ minima, a result that could allow, in principle,  to determine $q$.  Both the Luttinger parameter $\nu$ and the charge $q$ may  be also obtained from transport properties. For instance  the dc limit of the conductance of spinless fermions of charge $e$ and Luttinger parameter $K$  is given by\cite{kanefisher}
$
G_0= K e^2/h,
$
while  for $N$ chanels  it is given by\cite{affleck1}
\be
G_0=NK e^2/h =\nu \; \frac{(q e)^2}{h}.
\ee
We can infer from the latter relation that $\nu$ is the Luttinger
parameter associated with a single channel consisting into a bound-state of charge $qe$.  However we also see that the measurement of $G_0$ alone {\it does not } fully characterizes the Luttinger liquid state. To do so one needs an independent  measurement of the Luttinger parameter $\nu$. This could be achieved, in principle,  by the measurement  of the non linearities in the $I-V$ current-voltage curve in the presence of an impurity\cite{kanefisher, oreg, weiss, fendley}. 
%
%
\section{Generic bound-states and dynamical symmetry enlargement}
The requirement of the analycity of the correlation functions and the constraint of locality of the bound-states in the Luttinger liquid framework does not completely fix   the allowed charges of the possible bound-states: even if their charges $q$  are severly restricted  by the constraint (\ref{bssolution}),  there is still room for a fairly large number of possible bound-state solutions. It is obvious that at some point  the knowledge of the allowed charges $q$ and current $j$ quantum numbers should ultimately rely on the type of  ordering in  spin space  stabilized by the opening of the spin gap $\Delta$. At first glance it seems unlikely that more can be said about the possible
bound-states that can be stabilized by generic hamiltonians of the form (\ref{hamiltoniangeneric}). Fortunately it is  largely recognized that  for $generic$ interactions and fillings  the low-energy physics associated with multi-species interacting systems is captured by  Renormalization Group (RG) asymptotic trajectories which display an   enlarged symmetry\cite{lin, saleur, konik, boulat}. The so-called Dynamical Symmetry Enlargement (DSE) phenomenon. In the following we shall assume that such a DSE occurs. In the present case of $N$ fermionic species one may thus naturally expect that the $SU(N)$ symmetry of the non-interacting fermions might be dynamically enlarged at  low energies. But this is not the only  possibility. Other  dynamically enlarged symmetries are, as well, likely to occur. They
are  {\it dual} symmetries $\widehat{ SU(N)}$
obtained from $SU(N)$ by, in general non-local, duality transformations on the elementary fermions. Remarkably
enough,  the set of all possible dualities are known and fall
into a finite number af classes\cite{boulat}. As we shall show,   this will enable us to characterize a finite set of bound-state solutions of (\ref{bssolution}) that  we shall call {\it generic} bound-states. They are generic in the sense that, thanks to the DSE mechanism, they are the ones which are likely to be stabilized for a generic interaction. The strategy we shall
adopt in the following, will be to  use the fact that due to the presence of a spin gap $\Delta$,   the low-energy  wave functions has to be  singlets of  either the $SU(N)$ group or the $\widehat{ SU(N)}$ groups. The latter conditions, when translated in terms of the fermionic charges and currents ${ Q}_a$ and ${J}_a$, will yield    constraints on the total charge and current zero-mode operators ${ Q}$ and ${ J} $ and hence on the bound-state charge and current quantum numbers $(q, j)$. 

Assuming spin-charge separation and weak enough interactions,  
the low-energy physics of the generic hamiltonian (\ref{hamiltoniangeneric}) is to be described by the sum of two commuting charge and spin hamiltonians ${\cal  H} =  {\cal  H}_{} + {\cal  H}_s$ where ${\cal  H}_{}$ is given
by (\ref{luttinger}) and ${\cal  H}_s$ describes the spin fluctuations. In order to discuss the properties of ${\cal  H}_s$, and as we shall focus on the symmetry properties, it is  useful to describe the dynamics in the spin sector using non-abelian bosonization\cite{gogolin}. To this end let us introduce the right-left $SU(N)$ spin currents 
\be
  {\cal I}_{L(R)}^A =  \sum_{(a,b)=1}^N\psi^{\dagger}_{a, L(R)} T^A_{ab} \; \psi^{}_{b, L(R)},
\label{suncurrents}
\ee
where $T^A$, $A=(1,..., N^2-1)$, are the  generators of the Lie algebra of $SU(N)$ which are normalized as ${\rm Tr}(T^A T^B)= \delta^{AB}/2$. These currents satisfy  the $SU(N)_1$ Kac-Moody algebra given by the operator product expansion (OPE)
\be
{\cal I}_{L(R)}^A(x) {\cal I}_{L(R)}^B(y) \sim \frac{-\delta^{AB}}{8 \pi^2(x-y)^2} \pm \frac{f^{ABC}}{2\pi(x-y)} {\cal I}_{L(R)}^C(y).
\label{km}
\ee
In terms of these quantities the effective hamiltonian   in the spin sector may be written as a Wess-Zumino-Witten-Novikov (WZWN) $SU(N)_1$\cite{gogolin} perturbed by a marginal current-current interaction
\bea
{\cal H}_s &=& \frac{2\pi v_s}{N+1 } \int dx \; \sum_A [({\cal I}_L^A)^2 + ({\cal I}_R^A)^2]  +\int dx \; 
\sum_{AB} \; g_{AB} \;  {\cal I}_{ L }^{A}{\cal I}_{ R}^{B}. 
\label{hamiltonianspin}
\eea
When $g_{AB}=0$, the first part of the hamiltonian describes the  spin dynamics of $N$ free fermions with independent $SU(N)_L$ and $SU(N)_R$ symmetries. With these definitions, the statement of the DSE  phenomenon  can be phrased as follows: when the interaction is relevant the couplings $g_{AB}(t)$ grows with the RG-time $t$ and ultimatly reach some attractive ray where the symmetry is dynamically enlarged to some group $G$. 
As stated above  the symmetry can be maximally enlarged in the infrared to $G=SU(N)$ but as well to duals\cite{boulat} of $SU(N)$ in which cases $G=\widehat{ SU(N)}$. 
The constraints on the bound-states quantum numbers $(q, j)$ will be different. In the following
we shall assume that the symmetry is dynamically enlarged up to small symmetry breaking corrections.

%
%

\subsection{  $SU(N)$ Bound-States}
Let us start by discussing the simplest case of a maximally enlarged $SU(N)$ symmetry.  In this case the RG trajectory has the asymptotic $g_{AB}(t) \sim g(t) \delta^{AB}$ and the interacting part of (\ref{hamiltonianspin}) takes the asymptotic  $SU(N)$ invariant form
\be
{\cal H}_{\rm int} = g \int dx \;  \sum_A {\cal I}_{ L }^{A}{\cal I}_{ R}^{A}.
\label{hintsun}
\ee
When $g>0$ a spin gap opens and the ground state of (\ref{hintsun}) displays an (approximate) $SU(N)$ symmetry.
More precisely, the  effective  low-energy symmetry is  given by the diagonal group $SU(N)=SU(N)_L \times SU(N)_R|_{\rm diag}$   which  is  generated by
\be
{\cal I}_{  }^{A} = \int\; dx\; ({\cal I}_{ L }^{A} + {\cal I}_{ R }^{A}).
\label{sun}
\ee
Thanks to the gap in the $SU(N)$ sector the low energy sector is obtained by  projecting into the $SU(N)$ singlet sector 
\be
{\cal I}^A\equiv 0, \; \; \; A= 1,..., N^2-1.
\label{singletcurrent}
\ee
The latter equations    impose constraints  for the eingenvalues $Q_a$ and $J_a$  of the charge and current operators (\ref{QaJa}). Indeed let us consider the $N-1$ conserved charges associated with the $SU(N)$ symmetry. They are the subset of Cartan generators $h^{\alpha}$, $\alpha=(1,..., N-1)$, of the $SU(N)$  generators that are mutually commuting: $[h^{\alpha}, h^{\beta}] = 0$. They express in terms of the  fermions charges (\ref{QaJa}) (see the Appendix) as 
\be
h^{\alpha} = \sum_{a=1}^N \omega^{\alpha}_a { Q}_a,
\label{cartan}
\ee 
where the vectors $\vec \omega_a$ satisfy:
$\vec \omega_a \cdot \vec \omega_b =\delta_{ab} - 1/N $ and $\sum_{a=1}^N \vec \omega_a = 0$. Eq.(\ref{singletcurrent}) implies in particular $h^{\alpha} = 0$ for all $\alpha=(1,..., N-1)$ which together with the property $\sum_{a=1}^N  \vec \omega_a = 0$ yields for the species charges
$Q_a$:
\be
Q_a = n \in \mathbb{Z}, \;  a=(1,...,N).
\label{Q_aconstraint}
\ee
 As Eq.(\ref{singletcurrent}) does not yield other constraints on the values of the current quantum numbers than (\ref{periodic}),
we thus find for the total charge and current eigenvalues
\bea
&N& \;  {\rm even } \hspace{1.cm}  Q=n \; N, \; 
 J=2 m,  
\label{SUNqeven}
\\
 &N& \;  {\rm odd } \hspace{1.1cm}  Q=n \; N, \; J= m,\;  
\label{SUNqodd} \\
& & \, \hspace{1.9cm} (n  \pm m ) \; {\rm even},
\nonumber 
 \eea
where $(n,m)$ are relative integers. From (\ref{bosoqm},\ref{fermiqm})  we immediately find
\be
 (q = N, j=1).
\label{SUNbs}
\ee
The above solution satisfies the  constraint (\ref{bssolution})
and we identify these bound-states as charge $N$ bosons for $N$ even and charge $N$ fermions for $N$ odd, both with density $\rho_{\rm BS}=\bar \rho$ and a preserved Fermi momentum  $P_F=k_F$.  
%
%
\subsection{ Duals $\widehat{SU(N)}$ Bound-States}
On top of the dynamical  enlargement of the $SU(N)$ symmetry there are three other possibilities of DSE which are related to emergent duality symmetries\cite{boulat}. These dualities $\Omega$ act on one chiral sector of  the theory and in particular on the $SU(N)$ currents as follows:
\be
\Omega: {\cal I}^A_{L(R)} \rightarrow \widehat{{\cal I} }^A_{L(R)},
\label{duality}
\ee
where:
\be
 \widehat{ {\cal I}}^A_{L} = {\cal I}^A_{L}, \; \; \; \widehat{ {\cal I}}^A_{R} = \sum_B \; \Omega^A_B {\cal I}^B_{R},
\label{dualcurrents}
\ee
with $\Omega^2= 1$.  These dualities are symmetries of  the problem and preserve the Kac-Moody algebra (\ref{km}). Therefore,  to any set of initial conditions of the RG flow  $g_{AB}(0)$ that are attracted by the $SU(N)$ invariant ray there exists
models with couplings $\widehat{ g}_{AB}(0)=\sum_C \Omega_{A}^C g_{CB}(0)$ that will flow toward 
\be
{\cal H}_{\rm int} = \widehat{ g} \int dx \sum_A \widehat{ {\cal I}}_{ L }^{A}\widehat{ {\cal I}}_{ R}^{A}.
\label{hintsundual}
\ee
Similarly to the $SU(N)$ case,  when a spin gap opens and the ground state of (\ref{hintsundual})  displays an approximate  dual $\widehat{ SU(N)}$ symmetry  generated by
\be
\widehat{ {\cal I}}^A = \int dx \;  (   {\cal I}^A_L + \sum_B \Omega^A_B {\cal I}^B_R).
\label{dualsun}
\ee
We can   now look   at the constraints imposed on  the total charge and current quantum numbers $Q$ and $J$  when a spin gap is present.  The low energy sector we are interested with is the $\widehat{ SU(N)}$-singlet sector obtained by the projection
\be 
\widehat{ {\cal I}}^A \equiv 0.
\label{singletdual}
\ee
The resulting constraints on the charge and current operators ($Q$, $J$) can then be obtained from the knowledge of the duality $\Omega$ in (\ref{duality}).  Remarkably enough, the set of all possible $\Omega$ is known\cite{boulat}  and  fall into a finite number of classes  named ${\cal A}_{\mathbf{I}}, {\cal A}_{\mathbf{II}}$ and ${\cal A}_{\mathbf{III}}$. They  act  (up to a simultaneous  change of basis in the $L(R)$ chiral sectors) on the  left and right  fermions  (\ref{chiralfermions}) as 
\be
\Omega: \psi^{}_{a,L(R)} \rightarrow \widehat{\psi}^{}_{a,L(R)},
\ee
where
$
\widehat{\psi}^{}_{a,L} = \psi^{}_{a,L}
$
and
\bea
{\cal A}_{\mathbf{I}} &:& \widehat{\psi}^{}_{a,R}=\psi^{\dagger}_{a,R},  \\
\label{A1}
{\cal A}_{\mathbf{II}} &:& \widehat{\psi}^{}_{a,R} = \sum_{b=1}^N
 {\cal J}^{}_{ab} \psi^{\dagger}_{b,R} \;   \; ({N \; \rm  even}),  \\
\label{A2}
 {\cal A}_{\mathbf{III}} &:& \widehat{\psi}^{}_{a,R}= \sum_{b=1}^N ({\cal I}_{p})_{ab} \psi_{b,R}. 
\label{A3}
\eea
In the above equations the matrix
${\cal J}_{ab} = (-i \sigma_2) \otimes \mathbb{I}_{N/2}$ is the $SP(N)$ metric and 
${\cal I}_{p}$ ($0<p < N$) is the diagonal matrix with $N-p$ entries $+1$ and $p$ entries $-1$. In order to obtain   the  constraints on $Q=\sum_{a=1}^N Q_a$ and $J=\sum_{a=1}^N J_a$ imposed by the singlet dual projection (\ref{singletdual}) we use the fact that in the dual $\widehat{ SU(N)}$ ground state of (\ref{hintsundual}) the dual charge and current eigenvalues  $\widehat{ Q}$ and $\widehat{ J}$ satisfy (\ref{SUNqeven}) and (\ref{SUNqodd}).
With the knowledge of $\Omega$ in each class we then find
for both ${\cal A}_{\mathbf{I}}$ and ${\cal A}_{\mathbf{II}}$
classes
\bea
{\cal A}_{\mathbf{I}}:\; &N& \;  {\rm even } \hspace{1.cm}  Q=2 \; n, \; 
 J= m  N,
\label{A1dualityboson}
\\
 &N& \;  {\rm odd } \hspace{1.1cm}  Q=n \;  \; J= m N,
\label{A1dualityfermion} \\
& & \, \hspace{1.9cm} (n  \pm m ) \; {\rm even},
\nonumber 
 \eea
\bea
{\cal A}_{\mathbf{II}}:\; &N& \;  {\rm even } \hspace{1.cm}  Q=2  n, \; 
 J= m N.
\label{A2duality}
\eea
For  the class ${\cal A}_{\mathbf{III}}$ the constraints are the same 
as the ones given in the $SU(N)$ case by (\ref{SUNqeven}) and (\ref{SUNqodd}) and does not yield to a new  bound-state solution   but the $(q=N, j=1)$ one. As we shall see in the next section the duality class  ${\cal A}_{\mathbf{III}}$ provides for an internal structure of the bound-states. 
In contrast, both ${\cal A}_{\mathbf{I}}\; {\rm and} \;  {\cal A}_{\mathbf{II }}$ duality classes yield to new selection rules and hence to new types of bound-states. The main reason for this is that they contain the charge conjugation operator
\be 
\mathbb{C}: \psi^{}_{a, R} \rightarrow \psi^{\dagger}_{a, R},
\label{chargeconjug}
\ee
which induces an electromagnetic duality and exchanges the  charge and current operators in (\ref{SUNqeven}) and (\ref{SUNqodd}). The new  bound-states solutions   depend on the parity of $N$.

-  {\it $N$ even. } In this case the bound-states are bosons and for the two duality classes ${\cal A}_{\mathbf{I}}\; {\rm and} \;  {\cal A}_{\mathbf{II }}$ we find charge $q=2$ states with
\be
(q =2, j=N/2),
\label{dualboson}
\ee
which satisfy the locality constraint (\ref{bssolution}). As we shall see below, these bound-states correspond to genuine p-wave and s-wave  pairing states  for the duality classes ${\cal A}_{\mathbf{I}}\; {\rm and} \;  {\cal A}_{\mathbf{II }}$. Their density is from (\ref{bsdensity}) $\rho_{\rm BS}=N \bar \rho/2$ and their Fermi momentum is enlarged to $P_F=Nk_F/2$.

-  {\it $N$ odd.} In this case the ${\cal A}_{\mathbf{I}}$ class yields to a fermionic bound-state of  charge $q=1$ with
\be
(q =1, j=N),
\label{dualfermion}
\ee
satisfying (\ref{bssolution}). Though this bound-state has  the same charge as the elementary fermions,  it is  of a completely different nature. The bound-state density  in this case is $\rho_{\rm BS}=N \bar \rho$ and the fermionic bound-state  carries  left and right  momenta $\mp P_F$ with an enlarged Fermi momentum $P_F = Nk_F$. As we shall see, these states are local composite fermions.

Let us end this section by noticing that the two bound-state solutions, (\ref{dualboson}) and (\ref{dualfermion}) (associated with  the duality classes  ${\cal A}_{\mathbf{I}}\; {\rm and} \;  {\cal A}_{\mathbf{II }}$) are duals to the $SU(N)$ bound-states (\ref{SUNbs}).  The corresponding duality symmetry of the bound-sate Luttinger liquid hamiltonian, which is also a symmetry
of the bound-state equation (\ref{bssolution}), is given by
\bea
&q& \;  {\rm even } \hspace{1.cm}  (q, j) \rightarrow (2j, q/2),
\label{dualityboson}
\\
 &q& \;  {\rm odd } \;  \hspace{1.1cm}  (q, j) \rightarrow (j, q), 
\label{dualityfermion}
\eea
together with $\Phi_c \leftrightarrow \Theta_c$ and $K \leftrightarrow 1/K$. The above duality transformations 
preserve the bosonic and fermionic selection rules
(\ref{bosoqm}) and (\ref{fermiqm}) and hence map a bosonic (fermionic) Luttinger liquid to another bosonic (fermionic) Luttinger liquid.  The above duality transformations (\ref{dualityboson}) and (\ref{dualityfermion}) are equivalent, upon rescaling the fields in the dimensionless basis, to the one of a bosonic Luttinger liquid: $\bar \Phi \rightarrow \bar \Theta/2$,
$\bar \Theta  \rightarrow 2\bar \Phi$ and $\nu \rightarrow 1/4\nu$, and to that of a fermionic Luttinger liquid: $\bar \Phi \leftrightarrow \bar \Theta$ and $\nu \rightarrow 1/\nu$\cite{pham}. 
%
%
\section{ Wave Functions of the  $SU(N)$ and $\widehat{ SU(N)}$ bound-states}

So far we have determined the charge and current quantum numbers $(q, j)$ of the generic bound-states and it remains to characterize them in terms of the elementary fermions. The idea  is to look
at bare operators made of the lattice fermions $c_{a,i}$, in either $SU(N)$ or $\widehat{ SU(N)}$ singlets, which after averaging over the gapped spin degrees of freedom have a finite overlap with
the bound-state creation operators (\ref{bosonfermionexpansion}).
The choice of the bare wave function is of course not unique 
but, as we shall see, some choices turn out to be physically more transparent. In the case of  $SU(N)$ bound-states, the natural
choice is a local singlet with wave function made of the $N$ elementary fermions. For the $\widehat{ SU(N)}$ bound-states, the situation is less obvious since the  dual enlarged symmetries  generated by (\ref{dualsun})  are  {\it non local} in the original fermion basis (\ref{chiralfermions}). There exists though, for each class of duality,  subgroups $G_{\parallel}$ of  $\widehat{ SU(N)}$ that act locally in the elementary fermion basis. These are the subgroups of $SU(N)$ that are invariant under the duality transformations (\ref{A1}, \ref{A2}, \ref{A3}). The transformations $U\in G_{\parallel}$ act simultaneously on the two chiral sectors
\be
\psi^{}_{a, L(R)} \rightarrow U^{}_{ab} \psi^{}_{b, L(R)},
\ee
 with, for
\bea
{\cal A}_{\mathbf{I}} &:&   U \in O(N), \\
\label{UA1}
{\cal A}_{\mathbf{II}} &:&  U \in SP(N), \\
\label{UA2}
 {\cal A}_{\mathbf{III}} &:& U \in S(U(N-p) \times U(p)).
\label{UA3}
\eea
These latter local symmetries  $G_{ \parallel}$  will help us to characterize the bound-states stabilized in each duality class as 
local $G_{ \parallel}$-singlet operators made of the elementary  fermions.
%
%
\subsection{ $SU(N)$ Bound-States: Baryons }

Let us start with the simplest case of the $SU(N)$ singlets bound-states (\ref{SUNbs})
\be
(q=N, j=1),
\label{baryons}
\ee
and hence $P_F=k_F$. In terms of the elementary  electrons these excitations are naturally  related to a bound-state made of $N$ electrons in an $SU(N)$ singlet state
\be
B_{N,i}^{\dagger} =  c_{1,i}^{\dagger}...c_{N,i}^{\dagger},
\label{baryonN}
\ee
which is a boson for $N$ even and a fermion for $N$ odd. 
By analogy with  QCD, we may call these bound-states baryons.
Using the low-energy expansion (\ref{chiralfermions}) and averaging over the spin degrees of freedom in the $SU(N)$ ground state of (\ref{hintsun}) we find
\bea
\Psi^{\dagger}_{B/F} &=& <(B_N)_{}^{\dagger}(x) >|_{SU(N)} \sim  \sum_{\bar J=-N}^{N}  \alpha_{\bar J} \; \re^{i \bar J P_F x +i\sqrt{\pi} (\bar \Theta + \bar J \bar \Phi)},
\label{baryonsun}
\eea
where $B_N^{\dagger}(x)= B_{N,i}^{\dagger}/(a_0)^{N/2}$ and  the sum is over   $\bar J$ even for $N$ even  (bosons) and $\bar J$ odd for $N$ odd (fermions). Similar considerations lead for the relative   density operator,
\be
n_i = \sum_{a=1}^N c^{\dagger}_{a,i}c^{}_{a,i} - N \bar \rho \; ,
\label{density}
\ee
to 
\bea
 \rho(x) &=& < n(x)>|_{SU(N)} \sim  q \;  \partial_x \bar \Phi /\sqrt{\pi} + \beta_{2} \;  \re^{2 i P_F x + 2i \sqrt{\pi} \bar \Phi} + {\rm h.c.},
\label{densitysun}
\eea
with $n(x) = n_i/a_0$ and $q=N$. In both expressions (\ref{baryonsun}) and (\ref{densitysun}) we have rescaled the charge fields $\Phi_c$ and $\Theta_c$ according to (\ref{barphitheta}) with $q=N$ and $j=1$, i.e. $\bar \Phi = \Phi_c/\sqrt{N}, \bar \Theta= \Theta_c \sqrt{N}$. The coefficients $\alpha_{\bar J}$ and $\beta_{\bar J}$ are related to  the primary operators of the $SU(N)_1$ WZWN model  as
\bea
\alpha_{\bar J}&\simeq& (\gamma^*)^{(N-\bar J)/2}  < {\rm Tr}(\Phi^{(N-\bar J)/2)}]>|_{SU(N)},\; \alpha_{-\bar J}= \alpha_{\bar J}^* ,\nonumber  \\ 
\beta_{2}&\simeq&  < {\rm Tr}(\Phi^{(1)})>|_{SU(N)},  \;  \beta_{-2}= \beta_{2}^* ,
\label{alphabetasun}
\eea
where $\Phi^{(m)}$ is the primary operator of $SU(N)_1$ that transforms according to the representation of $SU(N)$ consisting into Young tableau with $m$ boxes and one column. Finally  $\gamma=\pm i$ is a cocycle (see the Appendix). The operator $\Phi^{(m)}$  has the scaling dimension $d_m= m(1-m/N)$ and therefore the coefficients
(\ref{alphabetasun}) scale with the spin gap $\Delta$ as
$\alpha_{\bar J} \sim \Delta^{(N^2 -\bar J^2)/4N}$ and  $\beta_2 \sim \Delta^{1-1/N}$. We thus find that, to this order,  the expressions for both (\ref{baryonsun}) and  (\ref{densitysun}) match  the expansions (\ref{bosonfermionexpansion}) and (\ref{densitybsboson}).
Notice though that  higher harmonics in $2mk_Fx$, $m>1$, are missing in the density operator expansion  (\ref{densitysun}). In any case, the coefficients $\alpha_{\bar J}$ and $\beta_{\bar J}$ are expected to be renormalized. Indeed, from the  renormalization group point of view,  the integration over the (high energy)  spin degrees of freedom is expected to generate corrections to (\ref{hintsun}). Among these, for instance,  are oscillating contributions to the hamiltonian density (\ref{hintsun}) like $\sum_m {\cal H}_m \re^{2imk_Fx}$ which renormalize  the various Fourrier components of both the  wave function (\ref{baryonsun}) and of the density (\ref{densitysun}) in which  higher components at $2mk_F x$ are thus expected to be generated. Another source of renormalization comes from the fact   that the $SU(N)$ symmetry of (\ref{hintsun}) is expected to be only approximate.   In general, there will be  subleading corrections due to symmetry breaking  operators which are supposed to be small for not too large anisotropies. 
 
At this point we may compare our findings with existing results.
The baryonic bound-states we have just described were 
found in the {\it attractive} $SU(N)$ Hubbard model, with Coulomb interaction $U < 0$, away from half-filling\cite{capponi}\cite{azaria2}\cite{ulbricht}.
DMRG results\cite{capponi} for  both $N=3$  and $N=4$ cases strongly support the existence of massless charge $q=3$ fermionic trions and  charge $q=4$ bosonic quartets bound-states  excitations,  while the  single fermions excitations are shown to be gapped.  In both cases, the baryon-baryon correlation
function exhibit  power-law behaviors with oscillations at wave vectors $\pm 2 k_F$. The physics in these cases were found to agree with that of spinless fermions or hard-core bosons\cite{giamarchi} in a wide range of densities $\bar \rho$ and couplings $U<0$. In particular,
for a sufficiently large $|U|$ and density, typically smaller than
$\bar \rho \sim 1/N$ (for which $\nu >1/2$ or $\nu > 1/\sqrt{3}$ for bosons and fermions respectively),  the baryon-baryon correlation function was found to be dominant. For larger $\bar \rho$ and smaller $|U|$,  the $2k_F$ density wave was found to be the dominant instability. All together these results  provide strong evidences for the relevance  of the bound-state description. Let us add that  further investigations also show that the effect of various anisotropies\cite{azaria2, roux2008}, like small breakings of the $SU(N)$ symmetry, does not modify the above picture. This shows that these baryonic bound-states are robust and generic and in particular  that the DSE hypothesis is sensible.
%
%
\subsection{ Dual $ {\cal A}_{\mathbf{I}}$ Bound-States }

As discussed above these bound-states are the duals
under (\ref{dualityboson}) and (\ref{dualityfermion}) of the
$SU(N)$ baryonic states. They are non-trivial states as they involve an enhanced Fermi momentum
$P_F$. The bound-states associated with the duality $ {\cal A}_{\mathbf{I}}$ are of two types depending on the parity of $N$ and are either charge $q=2$  bosons for $N$ even 
 (\ref{dualboson}) or charge $q=1$  fermions for  $N$ odd (\ref{dualfermion}). As the duality (\ref{A1}) is non local in terms of the elementary fermions, we shall look, as discussed above, at wave functions which are from (\ref{UA1}) $G_{\parallel}=O(N)$ singlets with either charge $q=2$ or $q=1$.
%
%
\subsubsection{$N$ even: P-Wave Pairing} 
The corresponding bound-state solution is given by (\ref{dualboson}) 
\be
(q=2, j=N/2)
\ee
and  hence $P_F=Nk_F/2$. Given the $O(N)$ symmetry of the problem it is natural to look at the p-wave symmetric lattice pairing operator
\be
\Pi^{\dagger}_i=\sum_{a=1}^N c^{\dagger}_{a,i}c^{\dagger}_{a,i+1}.
\label{pwave}
\ee
Using bosonization we find: $\Pi^{\dagger}_i/a_0 = \sin{(k_F a_0)} \; \Pi^{\dagger}(x)$ where
\be
 \Pi^{\dagger}(x) \simeq {\rm Tr}(\widehat \Phi^{(1)}) \; \re^{i\sqrt{4\pi/N}\;  \Theta_c}.
\label{pwavecont}
\ee
In the latter expression we have omitted terms that average to zero in the
$\widehat{ SU(N)}$ ground state of (\ref{hintsundual}).
The operator $\widehat \Phi^{(1)}$ entering in (\ref{pwavecont})
is the {\it dual} of the $SU(N)_1$ primary operator obtained
from $\Phi^{(1)}$ with help of the duality transformation 
$ {\cal A}_{\mathbf{I}}$ (see Appendix). It has the same scaling dimension $d_1=1-1/N$ and is odd under Parity, i.e. 
 ${\cal P} : {\rm Tr}(\widehat \Phi^{(1)}) \rightarrow - {\rm Tr}(\widehat \Phi^{(1)})$. As $\Theta_c \rightarrow \Theta_c$ under ${\cal P}$, we find that (\ref{pwavecont}) is odd as it should be.
A similar calculation yields for the density operator $n(x)$
(\ref{density})
\be
n(x) \simeq \sqrt{N/\pi}\;  \partial_x  \Phi_c, 
\label{densityA1}
\ee
were here again we have discarded terms that average to zero
in the $\widehat{ SU(N)}$ ground state of (\ref{hintsundual}).
The next step to be taken in order to obtain both the bound-state wave function $\Psi^{\dagger}_{B}(x)$ and the  density $\rho_{}(x)$ is to  average  over the spin degrees of freedom in the $\widehat{ SU(N)}$ ground state of (\ref{hintsundual}) with duality class  ${\cal A}_{\mathbf{I}}$. To do so we notice that,
as the duality transformations  ${\cal A}_{\mathbf{I, II, III}}$ are
symmetries of the problem, we have for any operator ${\cal O}$
\be
<\widehat{{\cal O}}>|_{\widehat{ SU(N)}} = <{\cal O}>|_{ SU(N)}.
\label{dualaverage}
\ee
where $\widehat{{\cal O}}$ is the dual of ${\cal O}$.
Hence we get 
\bea
\Psi^{\dagger}_{B}(x) &=& <\Pi^{\dagger}(x)>|_{\widehat{ SU(N)}} \sim  \alpha_0 \; \re^{i\sqrt{\pi} \bar \Theta} 
\label{pwaveA1boson}
\\
\rho_{}(x) &=& < n(x)>|_{\widehat{ SU(N)}}
\sim   q\;  \partial_x \bar \Phi/\sqrt{\pi},
\label{pwaveA1density}
\eea
with $q=2$ and
\bea
 \alpha_0 &\simeq&  <{\rm Tr}(\widehat \Phi^{(1)})>|_{\widehat{ SU(N)}}= <{\rm Tr}(\Phi^{(1)})>|_{SU(N)} \nonumber \\
&\sim& \gamma \;  \Delta^{1-1/N}.
\label{alpha_0A1}
\eea
In both Eqs.(\ref{pwaveA1boson}) and  (\ref{pwaveA1density})  we have rescaled the charge fields according to (\ref{barphitheta}) with $q=2$ and $j=N/2$ 
\be
\bar \Phi = \sqrt{N} \Phi_c /2, \bar \Theta = 2 \Theta_c/\sqrt{N}. \label{rescaleA1even}
\ee
Notice that in Eq.(\ref{alpha_0A1}) we have single out the  cocycle $\gamma$ to keep track of the parity transformation properties of the p-wave wave function, ${\cal P}: \gamma \rightarrow  \gamma^* = - \gamma$. In contrast with the baryonic bound-state wave function we find that
(\ref{pwaveA1boson}) and (\ref{pwaveA1density}) match  the expansions (\ref{bosonfermionexpansion}) and (\ref{densitybsboson}) only to leading order in the momentum expansion. In particular, the harmonics  at $\pm 2mP_F x$ with $P_F=Nk_F/2$,  are absent for both the bosonic wave function and the density.  This is not very satisfying as one of the hallmark of the bound-state solution in the dual ${\cal A}_{\mathbf{I}}$ class is the emergence of oscillations at the enlarged  wave vectors $\pm 2P_F=\pm Nk_F$.

{\it Composite Density.} As we shall now see, these harmonics are
generated by composite operators. Indeed, in the RG framework, we are at liberty to add to the effective hamiltonian any term which is compatible with the symmetries of the problem and that would be generated anyway at  energy scales $E  << \Delta$. In the following we shall accordingly  consider adding to the hamiltonian  (\ref{hintsundual}) the neutral (i.e. charge $Q=0$), ${\widehat{ SU(N)}}$-singlet and parity invariant  composite density operator with momentum components at $\pm N k_F $. To do this, let us first consider   the charge $Q=N$ and current $J=0$ singlet operator under  the $({ SU(N)_L} \times { SU(N)_R})|_{\rm diag}$ group. It is obtained from the  rank-N invariant tensor $\epsilon^{a_1...a_N}$ of ${ SU(N)}$ 
\bea
 \sum_{\{a_j\}} \epsilon^{a_1...a_N}
 \psi_{a_1, L}^{\dagger}...\psi_{a_{N/2}, L}^{\dagger}
\psi_{a_{N/2+1}, R}^{\dagger}...\psi_{a_N, R}^{\dagger}. \nonumber \\
\label{SUNprimaryinvariant}
\eea
The above operator is in fact proportional to the zero momentum
component of   the ${ SU(N)}$ baryon wave function (\ref{baryonsun}) when $N$ is even. We can now use the
duality transformation (\ref{A1}) to obtain the $ N k_F $
component of the  composite density operator, the
$- N k_F $ component being obtained with help of the parity
transformation. Using (\ref{A1})
and imposing parity invariance, we find for  the ${\widehat{ SU(N)}}$ composite density operator in the ${\cal A}_{\mathbf{I}}$ class ($N$ even) 
\bea
R_{N}(x) = \re^{iNk_Fx} \; R_{Nk_F}^{}(x) + \re^{-iNk_Fx} \; R_{-Nk_F}^{}(x)   \nonumber \\
\label{compositeA1}
\eea
where
\bea
R_{Nk_F}^{}(x) \simeq 
\re^{iN\pi/4} \sum_{\{a_j\}} \epsilon^{a_1...a_N}
 \psi_{a_1, L}^{\dagger}...\psi_{a_{N/2}, L}^{\dagger}
\psi_{a_{N/2+1}, R}^{}...\psi_{a_N, R}^{}.\nonumber \\
\label{compositedensityNkF}
\eea
 The phase factor $\re^{iN\pi/4}$ in (\ref{compositedensityNkF})
has been chosen in such a way that under ${\cal P}$, $R_{Nk_F}^{}(x) \rightarrow R_{Nk_F}^{\dagger}(x)= R_{-Nk_F}^{}(x)$. This 
ensures that  (\ref{compositeA1}) is indeed parity invariant.  Using  bosonization we  finally find
\bea
R_{N}(x)  
=  W_{N/2}(x)    \cos{( \sqrt{N \pi } \Phi_c + Nk_F x)}
\label{compositeA1even}
\eea
where 
\be
W_{N/2}(x) \simeq \gamma^{N/2}\; {\rm Tr}(\widehat \Phi^{(N/2)}).
\label{WNA1av}
\ee
In the above equation  $\widehat \Phi^{(N/2)}$ is the dual, under ${\cal A}_{\mathbf{I}}$,  of the $SU(N)_1$ primary operator $\Phi^{(N/2)}$ transforming in  the self-conjugate representation of $SU(N)$. 
As shown in the Appendix, the operator $W_{N/2}(x)$, which  has the scaling dimension $N/4$, is parity invariant and real: $W_{N/2}(x)={\cal P}W_{N/2}(x)= W^*_{N/2}(x)$.

 We may now write the contribution of $R_{N}(x)$ to the interacting hamiltonian (\ref{hintsundual})  as
\be
{\cal H}_{\rm int} \rightarrow {\cal H}_{\rm int} + \lambda \; \int dx \;  R_{N}(x), 
\ee
where $\lambda$ is some non universal coupling. For generic fillings, $k_F \neq 2\pi/N$,  it is  oscillating and  gives a negligible  contribution to the total hamiltonian. However, as discussed above, it does  have an effect on the renormalization of the different vertex operators. For instance, it generates the $\pm Nk_Fx$ components of both the p-wave bound-state wave function   and of the density operator (\ref{pwaveA1boson}) and (\ref{pwaveA1density}). To leading order in $\lambda$
we have
\bea
 \Psi^{\dagger}_{B}(x) &\rightarrow& \Psi^{\dagger}_{B}(x) + \lambda \; \delta \Psi^{\dagger}_{B}(x), \label{bosonshiftA1}
\\ 
\rho_{}(x) &\rightarrow& \rho_{}(x)+ \lambda\;  \delta \rho_{}(x), 
\label{deltapsirho}
\eea
where $\delta \Psi^{\dagger}_{B}(x)$ and $\delta \rho(x)_{}$ are given by the operator product expansions (OPE)
\bea
\delta \Psi^{\dagger}_{B}(x) &\sim&  <   R_{N}(z, \bar z)  \cdot \Pi^{\dagger}_{}(w,\bar w)>|_{\widehat{ SU(N)}} \nonumber \\
\delta \rho_{}(x)&\sim&  < R_{N}(z, \bar z) \cdot  n(w, \bar w)    >|_{\widehat{ SU(N)}} \nonumber \\
\label{deltapsirho}
\eea
where $z=\tau + i (x+a_0)$ and $w=\tau + i x$. Performing the necessary OPE 
and averaging over the spin degrees of freedom we find, 
for the bound-state wave function and the density operator,  the corrections to (\ref{pwaveA1boson}) and (\ref{pwaveA1density}) 
\bea
\Psi^{\dagger}_{B}(x) &\simeq&  \alpha_0 \; \re^{i\sqrt{\pi} \bar \Theta} + \alpha_2 \; \re^{i (  \sqrt{\pi} \bar \Theta + 2 \sqrt{ \pi } \bar \Phi +2P_F x)} +
\alpha_{-2} \; \re^{i (  \sqrt{\pi} \bar \Theta - 2 \sqrt{\pi } \bar \Phi - 2P_F x)} 
\label{pwavebosonA1} \\
\rho_{}(x) &\simeq& 
   q\;  \partial_x \bar \Phi/\sqrt{\pi} + \beta_2 \; \re^{i (   2 \sqrt{ \pi } \bar \Phi +2P_F x)} + \beta_{-2} \; \re^{-i (   2 \sqrt{ \pi } \bar \Phi +2P_F x)}
\label{pwavedensityA1}
\eea
where $\alpha_0$ is given by (\ref{alpha_0A1}) and\footnote{Notice the common factor $\gamma$ in the coefficients $\alpha_{\bar 0}$ and  $\alpha_{\pm 2}$ which  reflects the odd parity  of the  p-wave wave function.}
\bea
\alpha_2 &=& \alpha_{-2} \simeq \lambda\;  \gamma \Delta^{(N/4 - 1/N)} \nonumber \\
\beta_2&=&-\beta_{-2} \simeq  i\frac{a_0}{|a_0|}\; \lambda\;  \Delta^{N/4}.
\label{alphabetapm2A1}
\eea
In order to obtain the latter expressions we have made use of  (\ref{dualaverage}) and have  rescaled  the charge fields according to (\ref{rescaleA1even}). The momentum expansions  (\ref{pwavebosonA1}) and  (\ref{pwavedensityA1}) fit the general expressions (\ref{bosonfermionexpansion}) and (\ref{densitybsboson}) with non-vanishing coefficients up to $\pm 2P_F$. Higher momenta components  can be obtained similarly by including higher harmonics to the hamiltonian density or going  to higher order in $\lambda$. The important point is that these harmonics, being $\widehat{ SU(N)}$ symmetric, must carry multiples of $\pm Nk_F$. In this respect, the composite density $R_{N}(x)$ (\ref{compositeA1even}) is the minimal $\widehat{ SU(N)}$ invariant object that one can build from the bare fermions. As we shall see in the next section, it plays also a crucial role when discussing the incompressible phases associated with the dual phases. So far we have obtained the bound-state wave function assuming the dual symmetry
$\widehat{ SU(N)}$ is dynamically enlarged. Other corrections
to the coefficients $\alpha_{\bar J}$ and $\beta_{\bar J}$ are also
expected from small symmetry breaking operators. We  expect these corrections to be small and the p-wave bound-state
to be robust.   
%
%
\subsubsection{$N$ odd: Composite Fermions} 
We now discuss the bound-state solution (\ref{dualfermion})
\be
(q=1, j=N), 
\ee
which is a fermion with  an enlarged Fermi momentum $P_F=Nk_F$. As discussed above, this is a non trivial excitation since, though it has the same charge  than the elementary fermions, it carries an excess of current of $\pm (N-1)$ in its  left and right  components.  The situation is similar to the composite fermion construction\cite{wen, stonefisher, kane} (\ref{vertexbarl}). In the present case though the composite fermion can be made a {\it local} object thanks to the $N$  independent spin degrees of freedom. Let us consider for instance the fermionic charge $Q=1$ and  $O(N)$-symmetric lattice operator
\bea
\Xi_{i}^{\dagger}&=&  \sum_{\{a_j\}} \epsilon^{a_1...a_N}
 c_{a_1, i}^{\dagger}...c_{a_{\frac{N+1}{2}}, i}^{\dagger}
c_{a_{\frac{N+1}{2}+1}, i}...c_{a_N, i}^{}. \nonumber \\
\label{compositefermionlattice}
\eea
Using the low energy expansion (\ref{chiralfermions}) we  find that $\Xi^{\dagger}(x) = \Xi_{i}^{\dagger}/(a_0)^{N/2}$ has  left and right components at $\pm Nk_F$
\be
\Xi^{\dagger}(x) = \Xi^{\dagger}_{Nk_F} \re^{i Nk_F x} + \Xi^{\dagger}_{-Nk_F} \re^{-i Nk_F x},
\ee
where
\bea
\Xi^{\dagger}_{Nk_F} &\simeq& \sum_{\{a_j\}} \epsilon^{a_1...a_N} \psi_{a_1, L}^{\dagger}...\psi_{a_{(N+1)/2}, L}^{\dagger}
\psi_{a_{(N+1)/2+1}, R}^{}...\psi_{a_N, R}^{}, \nonumber \\
\Xi^{\dagger}_{-Nk_F} &\simeq& \sum_{\{a_j\}} \epsilon^{a_1...a_N} \psi_{a_1, R}^{\dagger}...\psi_{a_{(N+1)/2}, R}^{\dagger}
\psi_{a_{(N+1)/2+1}, L}^{}...\psi_{a_N, L}^{},
\label{compositefermionRL}
\eea
which, upon bosonization,  express as
\bea
\Xi^{\dagger}_{Nk_F} &\simeq& \gamma^{(N+1)/2}\; {\rm Tr}(\widehat \Phi^{(N + 1)/2})\;  \re^{i\sqrt{\pi}(\sqrt{N} \Phi_c + \Theta_c/\sqrt{N})}, \nonumber \\
\Xi^{\dagger}_{-Nk_F} &\simeq& \gamma^{(N-1)/2}\; {\rm Tr}(\widehat \Phi^{(N - 1)/2})\;  \re^{i\sqrt{\pi}(-\sqrt{N} \Phi_c + \Theta_c/\sqrt{N})}.
\eea
We may now  average over the spin degrees of freedom in the dual $\widehat{ SU(N)}$ ground state of (\ref{hintsundual}) with duality class  $ {\cal A}_{\mathbf{I}}$ to get the composite fermion
wave function
\bea
\Psi^{\dagger}_{F}(x) &=& <\Xi^{\dagger}(x)>|_{\widehat{ SU(N)}} \nonumber \\
& =&   \alpha_{1} \re^{iP_Fx} \; \re^{i\sqrt{\pi} (\bar \Theta +\bar \Phi)} + \alpha_{-1} \re^{-iP_Fx} \; \re^{i\sqrt{\pi} (\bar \Theta - \bar \Phi)},
\label{compositefermion}
\eea
with $P_F=Nk_F$ and 
$
 \alpha_{1} =\alpha_{-1} \simeq \Delta^{(N/4-1/4N)}
$.
In Eq.(\ref{compositefermion}) we have rescaled   the charge fields  according to  (\ref{barphitheta}) with $q=1$ and $j=N$
\be
 \bar \Phi = \sqrt{N} \Phi_c , \bar \Theta = \Theta_c/\sqrt{N}.
\label{rescalingA1odd}
\ee
The result (\ref{compositefermion}) shows that the local composite fermion (\ref{compositefermionlattice}) has a finite overlap with the bound-state solution (\ref{dualfermion}). In particular, when $\nu=1$, it can be interpreted as a free fermion with a sharp
extended Fermi surface with Fermi momentum $P_F$. Notice that this limit corresponds to strong {\it repulsive} interaction between the elementary fermions as $K=\nu/N=1/N$. 
The expression for the  density is the same as for the even
$N$ case (\ref{densityA1}), i.e. 
$
n(x) \simeq \sqrt{N/\pi}\;  \partial_x \Phi_c,
$
and there too,  the $\pm 2mP_F=\pm 2mNk_F$, $m >1$,  components are missing.  

{\it Composite density}. Following the same strategy as in the even $N$ case,  we are  led to consider adding to the interacting  hamiltonian (\ref{hintsundual})  the neutral, $\widehat{ SU(N)}$  symmetric and parity invariant composite density operator. In contrast with
the even $N$ case, when   $N$ is odd the latter operator must  have  momentum components at multiples of $\pm 2Nk_F $ since the quantum of current is now $j=N$.  The only
density operator 
with such a property is the self-dual,  i.e both 
${ SU(N)}$ and $\widehat{ SU(N)}$ symmetric, operator   given by
\be
R_{2N}(x) = \re^{2iNk_Fx} \; R_{2Nk_F}^{}(x) + \re^{-2iNk_Fx} \; R_{-2Nk_F}^{}(x) 
\ee
where
\bea
R_{2Nk_F}^{}(x)&\simeq&  
 \Pi_{j=a}^N\psi_{a, L}^{\dagger}\psi_{a, R}^{}, \nonumber \\
\label{compositedensity2NkF}
\eea
and  $R_{-2Nk_F}^{}   = R_{2Nk_F}^{\dagger}$. As (\ref{compositedensity2NkF}) is $SU(N)$
invariant, it is proportional to the identity operator which gives us
\be
R_{2Nk_F}^{}(x) \simeq  (\gamma)^N \; \re^{i\sqrt{4\pi N} \Phi_c}.
\label{R2NA1av}
\ee
As $N$ is odd and $\gamma^*=- \gamma$ we finally get
\be
R_{2N}(x) \simeq  (i\gamma) \sin{( \sqrt{4\pi N} \Phi_c + 2Nk_F x)}. 
\label{compositeA1odd}
\ee
At this point, it is worth stressing that  the  above expression is  $\cal P$ invariant despite the presence of the $\sin$ function.
This is due to the presence of the cocycle $\gamma$ since under
${\cal P}$, $\Phi_c \rightarrow -\Phi_c$ {\it and} $\gamma \rightarrow - \gamma$. As we shall see below,  this will be  of crucial importance when discussing boundary effects in the incompressible phase. Proceeding as with the p-wave wave function and performing the necessary OPE we find, using 
 (\ref{deltapsirho}) and (\ref{rescalingA1odd}), the expression for the  density
\be
\rho_{}(x)  \simeq   q\;  \partial_x \bar \Phi /\sqrt{\pi}\; + \beta_2\;  \re^{i (2 \sqrt{\pi} \bar \Phi + 2P_Fx)} + \beta_{-2}\; \re^{-i (2 \sqrt{\pi } \bar \Phi + 2P_F x)}
\label{compositedensity2PF}
\ee
with $q=1$, $P_F=Nk_F$ and
$
\beta_{2}= \beta_{-2} \sim (-i \gamma)\lambda a_0/|a_0|.
$
The momentum expansions  (\ref{compositefermion}) and  (\ref{compositedensity2PF}) match the general expressions (\ref{bosonfermionexpansion}) and (\ref{densitybsboson}) to leading non-trivial order with non-vanishing coefficients up to $\pm 2P_F$. In a similar way as for the p-wave bosonic bound-state, 
higher momenta components may be generated at higher orders in $\lambda$ and additional renormalizations of the coefficients
$\alpha_{\bar J}$ and $\beta_{\bar J}$ are to be expected. 
For the same reasons as for the p-wave bound-states, we expect
also,  despite the fact that the dual $\widehat{ SU(N)}$ symmetry of the $ {\cal A}_{\mathbf{I}}$ class is  only approximate, that the composite fermion will also be robust against small $\widehat{ SU(N)}$-symmetry breaking operators.

In sharp contrast with ${ SU(N)}$ baryonic bound-states, both the bosonic p-wave (\ref{pwave}) and the composite fermion (\ref{compositefermionlattice})  bound-states wave functions  display an enlarged Fermi momentum at $P_F=Nk_F/2$ and $P_F=Nk_F$. In order to account for this high momenta physics within the low-energy expansion (made around the two bare Fermi points $\pm k_F$),  we have seen that {\it composite operators} play a crucial role.  To start with, the composite fermion wave function itself is a bound-state  made of an elementary fermion and a composite of $(N-1)/2$ particle-hole excitations (\ref{compositefermionRL}) that account for excess of current needed to build up a total current $J=\pm N$.  In the bosonic case, we also find that the $\pm 2P_F=\pm Nk_F$ components of the bosonic wave function are due to the fusion with the composite density $R_N(x)$ (\ref{compositeA1even})  made of  $N/2$ particle-hole excitations. This is the signature that the ground state in the spin sector is highly non trivial. This is particularly true for the composite fermion since, as we see from (\ref{compositefermionlattice}),   there is no simple atomic limit
where this fermion can be defined contrarily with the baryonic
$SU(N)$ fermions (\ref{baryonsun}). To our knowledge, both the p-wave and composite fermion bound-states with $SO(N)$ symmetry have not  yet been predicted nor observed. 

%
%
\subsection{ Dual $ {\cal A}_{\mathbf{II}}$ Bound-States }
These bound-states exist for $N$ even only and correspond to the same bound-state solution as for the ${\cal A}_{\mathbf{I}}$ class
\be
(q=2, j= N/2),
\ee
and here again the Fermi momentum $P_F=Nk_F/2$ is enlarged.
Though the situation looks similar,  in the present case  the bound-state wave function and the underlying physics   is different. The main reason for
this is that the relevant local symmetry at present is $SP(N)$ and this has important consequences. These bound-states were studied in  Refs.(\cite{phle, capponi2, roux2008, phlezn}) in the context of cold fermionic
atoms with hyperfine spin $F=(N-1)/2$. In the following we shall review some of these previous findings in the light of the present work.

The $SP(N)$ symmetry  has two important consequences. First is the symmetry of the bound-state wave function which has to be a local $SP(N)$ singlet which implies an  s-wave pairing of the BCS type in contrast to the p-wave wave function of the class $ {\cal A}_{\mathbf{I}}$. A local lattice operator with this property is given by 
\be
P^{\dagger}_i = \sum_{a,b}\; c^{\dagger}_{a,i}c^{\dagger}_{b,i} {\cal J}_{ab}
\label{swave}
\ee
where  ${\cal J}_{ab}$ is the $SP(N)$ metric defined in (\ref{A1}).
To get a better understanding of the physics behinds (\ref{swave})
we may use a basis where the $N$ spin indices correspond to the 
$2F+1$ spin components of a half integer spin $F$: $a=(-F,...,F)$.
The $SP(N)$ metric is then proportional to the Clebsh-Jordan coefficient projecting onto the total spin-zero subspace:   
${\cal J}_{ab}\simeq <a, F; b,F |00>$. Hence (\ref{swave}) may be seen as  the s-wave  BCS wave-function for a half-integer spin $F$. 

The second consequence is the existence, on top of the $SP(N)$ symmetry,  of a discrete local  $\mathbb{Z}_{N/2}$ symmetry for $N>2$
\be
c^{\dagger}_{a,i} \rightarrow \re^{2im\pi/N} c^{\dagger}_{a,i}, \; m=0,..., N/2 - 1.
\label{zn}
\ee
 As discussed  in (\cite{phle}, \cite{phlezn}) the latter $\mathbb{Z}_{N/2}$ symmetry
plays a crucial role in the low-energy limit and the associated excitations are related to that of generalized  two-dimensional $\mathbb{Z}_{N/2}$ Ising models\cite{fateev}.  In a similar way as for the Ising model, these models  display a two-phase structure:  an  ordered phase where the $\mathbb{Z}_{N/2}$ is spontaneously broken, and a disordered phase where it is not. Accordingly there exist $N/2-1$, mutually non-local, order and disorder parameters $\sigma_k$ and $\mu_k$, $k=1,...,N/2-1$, such as in the ordered phase $<\sigma_k> \neq 0$ and $<\mu_k>$=0 and in the disordered phase $<\sigma_k> = 0$ and $<\mu_k>\neq 0$. These operators are of scaling dimensions $d_k= 2k(N-2k)/(N(N+4))$. On top of these spin fields, the $\mathbb{Z}_{N/2}$ CFT possesses neutral fields, $\epsilon_j \;  (j=1,...,[N/4])$, with scaling dimensions $d_j=4j(j+1)/(N+4)$ which are the thermal operators of the theory.

The important point with which we are concerned here is that the $\mathbb{Z}_{N/2}$ degrees of freedom have their own  energy scale, or gap $m$, which is independent of the
$SP(N)$ one $M$. In the generic situation $m\neq M$ and a faithful description of the physics involved in this system requires a detailed understanding of the interplay between both $\mathbb{Z}_{N/2}$ and $SP(N)$ degrees of freedom. This was  done in Ref.(\cite{phlezn}) using the CFT embedding $SU(N)_1 \sim SP(N)_1 \times \mathbb{Z}_{N/2}$ where the $\mathbb{Z}_{N/2}$ fluctuations are captured by the parafermionic CFT introduced in Ref.(\cite{fateev}). Without loss of generality,   we shall consider here the case where $M >> m$ and integrate out the $SP(N)$ degrees of freedom. The extension to the   $M\sim m$ can be done using the results of Ref.(\cite{phlezn}) and does not change qualitatively our results.

In the continuum limit the s-wave pairing operator (\ref{swave})
expresses in terms of the first order parameter $\sigma_1$
of the $\mathbb{Z}_{N/2}$ Ising model
\be
P^{\dagger}(x)\sim \sigma_1 \re^{i\sqrt{4\pi/N} \Theta_c},
\label{swpairingcont}
\ee
with $P^{\dagger}(x)=P^{\dagger}_i/a_0$, while the density operator is given only in terms of the charge field
\be
n(x)\sim \sqrt{N/\pi} \; \partial_x \Phi_c.
\ee
Given these results, two remarks are in order. Firstly  as (\ref{swave}) is parity invariant,   (\ref{swpairingcont}) has to be so. Therefore, as $\Theta_c$ is invariant under  ${\cal P}$,  $\sigma_1$ has to  also  be parity invariant which is indeed the case\cite{fateev}. Secondly, as the s-wave pairing term (\ref{swave}) is not invariant under the $\mathbb{Z}_{N/2}$ symmetry  (\ref{zn}),   the mere existence  of the bound-state (\ref{swave}) requires the $\mathbb{Z}_{N/2}$ symmetry to be spontaneously broken and hence the $\mathbb{Z}_{N/2}$ Ising model to be in its ordered phase with $<\sigma_1> \neq 0$.

{\it Composite Density}. As with the $ {\cal A}_{\mathbf{I}}$ class
of  bound-states, higher harmonics at $2m N k_F$ are missing to this order and have to be generated by some composite density operator. The relevant composite density operator in the present case  can be obtained following the strategy of the preceding
subsection  by taking the dual under  $ {\cal A}_{\mathbf{II}}$
of the charge $Q=N$ and $J=0$ $SU(N)$ singlet operator
(\ref{SUNprimaryinvariant}). Doing so, we find
\be
Q_{N}(x) =  \re^{iNk_Fx} \; Q_{Nk_F}^{}(x) + \re^{-iNk_Fx} \; Q_{-Nk_F}^{}(x)
\ee
where $Q_{\pm Nk_F}$ expresses in terms of the elementary fermions as
\bea
Q_{Nk_F}^{}(x)&\simeq&  \epsilon^{a_1...a_{N/2} b_1...b_{N/2}}
 \psi_{a_1, L}^{\dagger}...\psi_{a_{N/2}, L}^{\dagger}
\psi_{c_1, R}^{}...\psi_{c_{N/2}, R}^{} {\cal J}_{c_1 b_1}... {\cal J}_{c_{N/2} b_{N/2}},\nonumber \\
\label{compositedensityA2NkF}
\eea
and $Q_{- Nk_F}$ is obtained with the change $L \leftrightarrow R$. Using the results of Ref.(\cite{phlezn})
and averaging over the $SP(N)$ degrees of freedom we obtain in the limit $M>>m$
\be
Q_N(x) \sim \epsilon_1 \; \cos{(\sqrt{\pi N} \Phi_c + Nk_Fx)},
\label{compositeA2}
\ee
where $\epsilon_1$ is the first thermal operator of the $\mathbb{Z}_{N/2}$ Ising models which is even under ${\cal P}$. Using the OPE\cite{fateev} $ \epsilon_1(z, \bar z) . \sigma_1(w, \bar w)\sim \sigma_1(z, \bar z)$ we find, following the steps of the preceeding subsection,  the same expansion for both the bound-state  wave function $\Psi^{\dagger}_{B}$ and for the bound-state density $\rho$ as in  the p-wave case (\ref{pwavebosonA1}, \ref{pwavedensityA1}) with coefficients
\bea
 \alpha_0 &\simeq&   <\sigma_1> \sim m^{2(N-2)/N(N+4)}, \; 
\alpha_{\pm 2} \simeq \lambda \alpha_0,  \nonumber \\
\beta_{2}&=&-\beta_{-2} \sim i \lambda \frac{a_0}{|a_0|} <\epsilon_1> \sim i \lambda\; \frac{a_0}{|a_0|} m^{8/N(N+4)}.
\label{alpha_0A2}
\eea
The above findings are  in agreement with previous  results, obtained  by extended QMC and DMRG calculations, on a 1D lattice model with spin $3/2$ fermions ($N=4$)\cite{capponi2, roux2008}: at quarter filling ($\bar \rho =1/4$) an extended phase with deconfined s-wave BCS pairs (\ref{swave}) together with gapped single particle excitations was
shown to exist. In addition, density fluctuations with wave vector $2P_F=\pi$ were clearly observed in a wide range of parameters, a result which is consistent with a bound-state density $\rho_{\rm BS}=1/2$ when  $\bar \rho =1/4$,  $N=4$ and $q=2$.

%
%

\subsection{ Dual $ {\cal A}_{\mathbf{III}}$ Bound-States }
These are the last types of generic bound-states. As discussed previously, they have the same quantum numbers as the baryonic states with $SU(N)$ symmetry (\ref{baryons})
\be
(q=N, j=1)
\ee
and $P_F=k_F$. However, the duality (\ref{A3}) is still non-trivial and provides for an internal structure of the bound-states. This is the manifestation of the fact that  the local  symmetry group associated with the duality  class $ {\cal A}_{\mathbf{III}}$  is not $SU(N)$ but rather, from (\ref{UA3}),  $G_{\parallel}= S(U(p)\times U(N-p))$. Therefore,   one may naturally  anticipate that they are made of a bound-state of both $SU(p)$-singlet and $SU(N-p)$-singlet baryons. In order to shed light on the physics that hides behind (\ref{A3}), it is useful to first  consider the density per spin or species $n_a(x) = \rho_{a,i}/a_0$
\be
n_a(x) = \partial_x \phi_a/\sqrt{\pi} + (\Phi^1)_{a,a} \; \re^{(2ik_F x + i \sqrt{4\pi/N} \Phi_c)} +h.c.
\ee
where $(\Phi^1)_{a,a}$ are the diagonal components of the $SU(N)_1$ primary operator transforming in the fundamental representation of $SU(N)$. When averaging over the spin degrees of freedom in the $\widehat{ SU(N)}$ ground state of (\ref{hintsundual}) with duality class  $ {\cal A}_{\mathbf{III}}$
we make use of (\ref{dualaverage}) with 
\be
<(\Phi^1)_{a,a}>|_{\widehat{ SU(N)}} = <(\Phi^1 {\cal I}_p)_{a,a}>|_{{ SU(N)}}
\ee
where ${\cal I}_p$ is the diagonal matrix defining the duality
transformation (\ref{A3}). Using $SU(N)$ invariance we find
\bea
<n_a(x)>|_{\widehat{ SU(N)}} &=& \partial_x \Phi_c /\sqrt{N\pi}
+ (\beta_2 \re^{(2ik_F x + i \sqrt{4\pi/N} \Phi_c)} +h.c.),\; a=(1,...,p), \nonumber \\
&=& \partial_x \Phi_c /\sqrt{N\pi}
- (\beta_2 \re^{(2ik_F x + i \sqrt{4\pi/N} \Phi_c)} +h.c.),\; a=(p+1,...,N),\nonumber \\
\label{densitiedA3}
\eea
where 
$
\beta_2 \simeq \gamma \Delta^{1-1/N}.
$
The latter result shows that the $2k_F$ components of the density
waves of the $p$ species or spins, labeled $a=(1,...,p)$, are out of phase from those of the remaining $N-p$ ones, labeled $a=(p+1,...,N)$. Therefore,  the two density profiles are shifted by a distance 
\be
x_0=\pi/2k_F.
\label{x0}
\ee
Considering now the total density
\be
\rho(x) =\sum_a <n_a(x)>|_{\widehat{ SU(N)}},
\ee
we find, upon rescaling the charge fields, the same expansions as in the $SU(N)$ baryonic case (\ref{densitysun}) with coefficients at $\pm 2k_F$
\be
\beta_{2} \simeq (2p-N) \gamma \Delta^{1-1/N}, \; \beta_{-2} = - \beta_{2}.
\label{betap}
\ee
We notice that these coefficients vanish  when $p=N/2$ ($N$ even) due to the $\pi$ phase-shift. This effect is not expected to survive corrections due to symmetry breaking operators unless the system possesses an additional $\mathbb{Z}_{2}$ symmetry
interchanging the two sets $a=(1,...,p)$ and $a=(p+1,...,N)$. In any case, one may also define a relative density between the two sets, which reads (in an obvious notation) $\delta \rho(x) = \rho_p(x) - \rho_{N-p}(x)$, that exhibits $\pm 2 k_F$ oscillations.

From the above discussion we are  naturally led to look after a bound-state made of two $SU(p)$ and $SU(N-p)$ singlets separated by a distance $x_0$. With the notation of (\ref{baryonsun})) let us  consider now the wave function
\be
(B^p_N)^{\dagger}(x) = B_{p }^{\dagger}(x) B_{N-p }^{\dagger}(x+x_0).
\label{baryonA3p}
\ee
 For not too small densities $\bar \rho$, in which case $x_0$ is of order the lattice spacing $a_0$, one may use the low-energy expansion (\ref{chiralfermions}) and average over the spin degrees of freedom in the ${\widehat{ SU(N)}}$ ground state of class $ {\cal A}_{\mathbf{III}}$. As result we   find for the bound-state wave function
\be
(\Psi^p)^{\dagger}_{B/F} = <(B^p_N)^{\dagger}(x) >|_{\widehat{ SU(N)}},
\label{baryonA3}
\ee
the same expansion (\ref{baryonsun}) as for the $SU(N)$ baryons with, up to a phase,  the same coefficients $\alpha^p_{\bar J} =\alpha^N_{\bar J}
$. We notice at this point that one could also have defined the bound-state (\ref{baryonA3p}) at another value of the relative distance, $y \neq x_0 = \pi/2k_F$,  between the two $SU(p)$ and $SU(N-p)$ baryons in (\ref{baryonA3p}). In general, the corresponding amplitudes $\alpha^p_{\bar J}(y)$ are non-zero but  the $|\alpha^p_{\bar J}(y)|$ are   maximal at  $y= \pm x_0$, a result which is consistent with the behavior of  the density waves (\ref{densitiedA3}). The baryonic wave function  (\ref{baryonA3p}) or (\ref{baryonA3}) might be even or odd under the reflexion $x_0 \rightarrow - x_0$ as
\be
 B_{p }^{\dagger}(x) B_{N-p }^{\dagger}(x-x_0)= (-1)^{N-p} \; B_{p }^{\dagger}(x) B_{N-p }^{\dagger}(x+x_0).
\label{paritybaryonA3}
\ee
With these results at hand one may now draw the following  physical picture: the bound-states (\ref{baryonA3p}) may be seen as  symmetric or anti-symmetric pairs of baryonic $SU(p)$ and $SU(N-p)$ singlets. These pairs might be bosons or fermions depending on the parity of $N$. In the fermionic case, i.e. when $N$ is odd, the pair is made of a boson and a fermion. When $N$ is even the pair is bosonic and may consist of two charged $p$  and $N-p$  bosons ($p$ even) with a symmetric wave function, or  fermions ($p$ odd) with an antisymmetric wave function.  

Until now we made the assumption that the density per spin $\bar \rho$  is not too small so that $x_0 \sim 1/2\bar \rho$ is of order of the lattice spacing. When $\bar \rho << 1$ (which corresponds to the strong interaction regime)  we might expect the two  $SU(p)$ and $SU(N-p)$ singlets to be weakly bounded over a separation $\delta x$ such as $k_F \delta x <<1$. Although we have no general proof,  in this regime,  we expect the pairs to be unstable toward decoupling ,  for example due to  a  {\it repulsive} interaction between the two $SU(p)$ and $SU(N-p)$  baryons. This is actually what has been demonstrated\cite{azaria2} in the simplest case of $N=3$ and $p=2$ where, at small enough densities, a trionic bound-state made of a $F=1/2$ BCS pair and a single fermion was found to be unstable toward decoupling upon switching on a small repulsive interaction between them. It is  beyond the scope of the present work to elaborate on  the general case.

%
%

\section{Incompressible phases}
In the preceding sections we have provided for a description
of the low-energy physics of generic hamiltonians of the type (\ref{hamiltoniangeneric}) in terms of the bound-state that are stabilized by the opening of a spin gap $\Delta$. Once a bound-state solution of (\ref{bssolution}) is given in terms of $(q, j)$,
the low-energy physics at energy scales much smaller than
the spin gap is captured by a Luttinger liquid hamiltonian with momentum scale $P_F= j k_F$ ($j=N/q$). Equipped with this result, it is  natural to look at the possible instabilities of such  a state in the regime $E << \Delta$. In the charge sector of the theory the most important instability is due to commensurability effects with the lattice and the opening of a Mott gap stabilizing an incompressible phase. In the following we shall  relate the nature of the Mott phases to that of the low energy bound-states we discussed above.

As is well known, the general strategy to investigate the Mott transition is to look at small umklapp perturbations to the Luttinger liquid state. In the framework of the bound-state Luttinger liquid we can express things in terms of the "dimensionless" charge fields $\bar \phi$ and $\bar \theta$
provided one uses  the bound-state density $\rho_{{\rm BS}}$ as 
the relevant parameter that controls the commensuration effects. 
To this end, we  shall   consider small perturbations of the Luttinger liquid hamiltonian 
\be
 {\cal H} \rightarrow {\cal H} + V_{\rm Mott},
\ee 
where ${\cal H}$ is given in (\ref{luttingerbar}) and $V_{\rm Mott}$ is any potential allowed by the symmetries of the problem which are, on top of charge conservation,  translational and parity invariance. Decomposing $V_{\rm Mott}$ in the basis of the vertex operators (\ref{vertexbar}) and taking into account the global $U(1)$ symmetry associated with charge conservation,  one finds that the allowed vertex operators lie in the zero charge sector $\bar Q =0$ and hence, carry even currents $\bar J= 2m$.
One thus has
\be
V_{\rm Mott}=  \sum_{m\ge 0} \lambda_m \int \; dx \; \re^{-2im\sqrt{\pi} \bar \phi(x)} + h.c. .
\label{mottpotential}
\ee
The constraint imposed by translational invariance on the lattice arises, after noticing that each term in the sum (\ref{mottpotential}) carries a momentum $P_m= 2m P_F$ ($P_F=\pi \rho_{\rm BS}$), from  the conservation of momentum   up to a lattice reciprocal vector $\equiv 2n \pi$. This   imposes the commensurability condition $2m P_F = 2n \pi$ which reads  in terms of the bound-state density
\be
 \rho_{\rm BS} = \frac{n}{m},
\label{commcond}
\ee
or in terms of the bare density
\be
\bar \rho = \frac{n}{m} \frac{1}{j  }.
\label{commcondbare}
\ee
 Keeping  the most relevant term in the expansion
(\ref{mottpotential}) compatible with the commensurability condition (\ref{commcond}), we are led to write the effective hamiltonian describing the Mott transition for commensurate  bound-state fillings as
\be
{\cal H} =   \int dx \; \left[\frac{u}{2} [\frac{1}{\nu} (\partial_x \bar \phi)^2 +
\nu (\partial_x \bar \theta)^2] +  \lambda \cos{(2m\sqrt{\pi} \bar \phi(x) + \eta)} \right],
\label{sinegordon}
\ee
where $\lambda$ is a non-universal  coupling and $\eta$ is a phase.
The last constraint on (\ref{mottpotential}) comes from the parity symmetry,  ${\cal P}: V_{\rm Mott} \rightarrow  V_{\rm Mott}$,  which should fix the phase $\eta$. The latter   depends on how the vertex operators $\re^{-2im\sqrt{\pi} \bar \phi(x)}$ tranforms under parity which  we find  a non-trivial issue for a  general bound-state Luttinger liquid. We  shall come back later to this problem when focussing on the particular cases of integer bound-state densities where $\eta$ plays a crucial role.

{\it Fractional Bound-State Fillings}. Let us first focus on   generic fractional fillings, i.e. when $(n,m)$ are co-prime integers.  The physics behind (\ref{sinegordon}) is well known\cite{giamarchi}. When $\nu m^2 \le 2$ the cosine term becomes relevant, a gap opens in the charge sector, and the system becomes an insulator. Translational
symmetry on the lattice which reads in term of the bosonic field
\be
\bar \phi(x) \rightarrow \bar \phi(x) +  P_{F}/\sqrt{\pi}
\label{translationPF}
\ee
is spontaneously broken leading to an   $m$-fold degenerated ground state.   The gapped elementary  excitations are solitons or kinks  that interpolate between two ground states and have a fractional charge
\be
Q_s= \frac{q}{m}.
\label{solitoncharge}
\ee
What we just described  is similar to what happens in a one species problem provided one uses as   the relevant physical  quantity the bound-state density $\rho_{\rm BS} =  \bar \rho j/q$ rather than the  species density $\bar \rho$. The fact that it is $\rho_{\rm BS}$ and not $\bar \rho$  that controls commensurability effects with the lattice has  an important consequence for integer bound-state densities and leads to new physics.

{\it Integer Bound-State Fillings}. 
Let us  now consider the case of integer bound-state densities:
\be
 \rho_{\rm BS}= n
\label{integerdensity}
\ee
which implies $m=1$ in (\ref{commcond}, \ref{sinegordon}). Since 
$\rho_{\rm BS}= j \bar \rho $ such a situation can only occur
when $j > 1$ (this is due to Pauli principle that requires $\bar \rho < 1$). This situation can therefore  only happen for bound-state solutions $(q,j)$ where the Fermi momentum is enhanced, i.e. $P_F =j k_F> k_F$. This is only possible when the number of species $N > 2$. In these cases, translation symmetry (\ref{translationPF}) remains  unbroken in the insulating or Mott phase and the ground state is not degenerate. This opens the interesting possibility that some of these insulators may be topological insulators. 
%
%
\subsection{Charge Edge States and Generic Bound-States}

The topological character of these insulating phases
rely on the possible existence of zero-energy modes (ZEM), or  edge states, in the problem\cite{kanehasan}. At the level of this work, where we focus on the instability of the bound-state Luttinger liquid, one may only address the possible existence of ZEM in the charge sector and can gain no information of what happens in the spin sector. Even in this case the situation is complex since, as we shall see, the existence of charge ZEM ultimately relies on the phase
$\eta$ in (\ref{sinegordon}) and hence on the way  the vertex operators transform under the parity symmetry ${\cal P}$. For a general bound-state Luttinger liquid, as said above, we find it a difficult problem. However, this issue can be solved for the generic
bound-state solutions we have discussed in the previous section.
Out of the five types of  generic bound-states, the constraint of an integer bound-state density (\ref{integerdensity}) can be possibly realized only with the classes ${\cal A}_{\mathbf{I}}$ and $ {\cal A}_{\mathbf{II}}$ for which $(q=2, j = N/2)$ or $(q=1, j=N)$. The relevant Mott potentials  in these cases are given by the {\it composite densities} $R_N(x)$, $R_{2N}(x)$ and $Q_{N}(x)$ of Eqs.(\ref{compositeA1even}),  (\ref{compositeA1odd}) and (\ref{compositeA2}). At integer bound-state density  these composite fields are not oscillating, charge neutral and parity invariant.  On top of that, as they all carry momentum $\pm 2P_F$, they  identify with the operators with the smallest scaling dimension in (\ref{mottpotential}). One then finds for $N$ even in both
${\cal A}_{\mathbf{I}}$ and $ {\cal A}_{\mathbf{II}}$ classes
\be
V_{\rm Mott} \simeq \int dx\; <R_N(x)>, \int dx\; <Q_N(x)>,
\ee
and for $N$ odd in the ${\cal A}_{\mathbf{I}}$ class
\be
V_{\rm Mott} \simeq \int dx\; <R_{2N}(x)>,
\ee
where  $<...>$ denotes the  average over the spin degrees of freedom in the corresponding dual ground states.  Upon rescaling the charge fields according to the "dimensionless" basis (\ref{barphitheta}) we finally end up with  two different types
of effective hamiltonians
\bea
{\cal H}_{B}&=& {\cal H}_{} + g\; \int dx \; \cos{(2\sqrt{\pi} \bar \phi)}, 
\label{hamiltonianB}
\\
 {\cal H}_{F}&=& {\cal H}_{} + i \gamma\;  g\; \int dx \; \sin{(2\sqrt{\pi} \bar \phi)},
\label{hamiltonianF}
\eea
where  ${\cal H}_{B}$ is the effective hamiltonian for {\it bosonic} p-wave or s-wave charged $q=2$ pairs for the classes ${\cal A}_{\mathbf{I}}$ and $ {\cal A}_{\mathbf{II}}$ and the  hamiltonian ${\cal H}_{F}$ describes  the charge $q=1$ composite fermions
of class ${\cal A}_{\mathbf{I}}$. The coupling constant $g$ is, from Eqs.(\ref{WNA1av}, \ref{compositeA2}, \ref{R2NA1av}), proportional to $<W_N(x)> \sim \Delta^{N/4}, <\epsilon_1> \sim m^{8/N(N+4)}$ for the bosonic bound-states
of class ${\cal A}_{\mathbf{I}}$ and $ {\cal A}_{\mathbf{II}}$ and for composite fermion bound-state of class ${\cal A}_{\mathbf{I}}$,  $g\simeq cst$. The two Mott potentials, in the bosonic and fermionic cases, are different as they involve $\cos$ and $\sin$ functions of the charge field  $\bar \phi$. They essentially differ in the way the parity symmetry ${\cal P}$ is realized. As under ${\cal P}$: $\bar \phi \rightarrow - \bar \phi$  {\it and} $\gamma \rightarrow \gamma^* = - \gamma$ both Mott terms are two independents ${\cal P}$-invariant potentials. As far as bulk properties are concerned, this difference has no important consequences but when investigating {\it boundary properties}, as the existence of possible edge states, it is crucial.

Let us now consider the system in the semi-infinite geometry $[0, \infty[$ with Open Boundary Condition (OBC) at $x=0$.  To get some insights   let  us  focus on  the Luther-Emery point at which the Luttinger parameter $\nu = 1$ and both hamiltonians (\ref{hamiltonianB}) and (\ref{hamiltonianF}) can be expressed in terms of  that of free massive fermions. Indeed introducing the chiral fermionic operators
\be
\Psi_{R(L)} \simeq  \exp{-i\sqrt{\pi}(\bar \theta \mp \bar \phi)}
\label{Chiralfermions}
\ee 
one may rewrite (\ref{hamiltonianB}) and (\ref{hamiltonianF})  as a 1D Dirac hamiltonian
\bea
{\cal H}_{B(F)}&=& -\int_0^{\infty} dx \; \Psi^{\dagger}h_{B(F) }\Psi, \\
h_{B(F) }&=&  i\sigma_3 \partial_x + m   \sigma_{2(1)},
\label{diracBF}
\eea
where $\sigma_{a=1,2,3}$ are the Pauli matrices, $\Psi$ is the two-component spinor
\be
\Psi= \left(\begin{array}{c} \Psi_{R} \\ \Psi_{L} \end{array}\right)
\label{spinor}
\ee
and $m= - \pi g$ is a mass parameter\footnote{Notice that the mass parameter do not depends on the cocycle $\gamma$ anymore. In the composite fermion case it is reabsorbed in the fermions $\Psi_{L(R)}$. For bosonic bound-states,  the $\cos$ term yields a cocycle $\gamma$ upon refermionization which, as shown in Ref.\cite{phleorignac}, is fixed to $\gamma=-i$ by the OBC. }. The fermionic operators
(\ref{Chiralfermions}) have different physical origins in both the
bosonic and fermionic cases. While in the composite fermion case 
described by ${\cal H}_{F}$, (\ref{Chiralfermions}) are allowed eigenstates of the Luttinger liquid, in  the bosonic case described by ${\cal H}_{B}$ they are not. In this case, the fermions (\ref{Chiralfermions}) are rather Laughlin quasi-particles (at $\nu =1$) states\cite{pham} that span the zero-charge sector of the Luttinger liquid spectrum  and always occur in particle-hole pairs.

The 1D Dirac hamiltonian possesses (for suitable boundary condition) a zero-energy  solution $\Psi_0(x)$  localized at the boundary\cite{su} at $x=0$. As  the mass terms in (\ref{diracBF}) differ for both bosonic and composite fermionic bound-states the localized ZEM wave function are different in both cases.
For the composite fermion bound-states we have
\be
\Psi_{0F}(x)= \sqrt{|m|} \; \re^{-|m|x} \left(\begin{array}{c} 1 \\ \ri\;  {\rm sgn}(m) \end{array}\right)
\label{zemfermion}
\ee
while for bosonic bound-states
\be
\Psi_{0B}(x)= \sqrt{|m|} \; \re^{-|m|x} \left(\begin{array}{c} 1 \\  -{\rm sgn}(m) \end{array}\right).
\label{zemboson}
\ee
The question is now whether such states exist for the lattice model
with open boundary conditions. To see this, let us as usual, modelize  the open boundary condition on the lattice  by
$
c_{a, i=0}=0, \; a=(1,...,N),
$
which implies for  the continuum fermions
\be
\Psi_{L,a}(0) +  \Psi_{R,a}(0) = 0
\ee
for each species.
We immediately find the corresponding boundary conditions on the rescaled fields
\be
\bar \phi(0) =  \frac{N\sqrt{\pi}}{2q},
\ee
and hence on the Dirac spinors for both the charge $q=1$ composite fermion and charge $q=2$ bosonic s-wave or p-wave bound-states
\bea
q=1 &: & \Psi_{R}(0)=-  \Psi_{L}(0) 
\label{fermionicobc}\\
q=2 &: & \Psi_{R}(0)= \re^{-iN\pi/2} \Psi_{L}(0).
\label{bosonicobc}
\eea
We now arrive at the important conclusion that { \it in the composite fermion case there is no massless edge state} localized at the boundary $x=0$ since the ZEM solution of the Dirac equation (\ref{zemfermion}) does not match with the OBC (\ref{fermionicobc}). In contrast, for bosonic bound-states, a ZEM solution may exist depending on the sign of the mass $m$. From (\ref{zemboson}) one finds that  for $N/2$ even and $m < 0$ and for $N/2$ odd and $m >  0$ the OBC on the lattice (\ref{bosonicobc}) is compatible with  (\ref{zemboson}). When such a ZEM exists, the ground state is to be doubly degenerated corresponding to the presence a fractional charge\cite{jackiw2} 
$\pm Q_{\rm edge}$ at the edge. As the bound-state charge is
$q=2$ the charge at the edge is
\be
Q_{\rm edge} = q/2 =1
\ee
in units of the elementary fermion charge. In a system with
open boundary conditions at both ends $[0, L]$, where $L$ is the system size, we may expect, by symmetry, a four-fold degeneracy (to the $\re^{-|m| L}$ accuracy). Thus, as far as the charge degrees of freedom are concerned, we   find a link  between the nature of the low energy bound-states and the possible topological nature of the associated insulating phases. Although all the discussion on the existence of  the ZEM  was made at the Luther-Emery point, we expect  our results to hold qualitatively  when one departs from $\nu=1$. The reason for this is that a value of  $\nu \neq 1$ reflects    the interaction between the fermions which, we expect, only affect the bulk properties\footnote{There is though a particular value of $\nu=2$ where the Mott potential displays an $SU(2)$ symmetry in the charge sector. In this case, as argued in Ref.(\cite{nonne1}), the number of edge states may be greater than two.}. With this said  two remarks are in order:

Firstly the fact that a given model exhibits either p-wave or s-wave pairings of the type ${\cal A}_{\mathbf{I}}$ or $ {\cal A}_{\mathbf{II}}$ is not a sufficient condition for it to become a topological insulator at integer   bound-state density; at issue is the sign of the mass term in (\ref{diracBF}) which is model dependent. On top of that, even though  one might expect  the charged edge states to be protected by the extended  dual ${\widehat{ SU(N)}}$ symmetries present in both classes ${\cal A}_{\mathbf{I}}$ and $ {\cal A}_{\mathbf{II}}$, we have no proof that they are stable\cite{pollmann}. 

 Secondly the above discussion focuses only on edge modes in the charge sector. It could be well that edge modes exist in the spin sector so that the above analysis does {\it not} allow us to conclude about the total degeneracy of the ground state. 
In particular,  the absence of edge states in the charge sector does not imply that a given system  is not a topological insulator. This is particularly true for the composite fermion case. 
Though we certainly believe that there are no charge edge states in these systems, there exists the possibility that  ZEM in the {\it spin sector} may be stabilized in the Mott phase.  In this respect, in the simplest case of $N=3$ at the filling $\bar \rho =1/3$, preliminary investigations\cite{azariaberg} on a particular $SO(3)$ invariant fermionic model that  display composite fermions as low energy excitations  may exhibit   spin-$1/2$ ZEM at each end of an  open chain. We hope to come back soon to this topic in a forthcoming publication\cite{azariaberg}.

The existence of charge edge states was first predicted in a one dimensional lattice bosonic system with extended interactions\cite{bergtorre}. In a very nice series of works, H. Nonne and co-workers\cite{nonne1, nonne2}, further demonstrated the existence of charge edge states in a system of spin $F=3/2$ fermions with $SP(4)$ symmetry at half-filling, i.e. $\bar \rho=1/2$.  In these phases, called Haldane Insulator\cite{bergtorre}, the charge degrees of freedom are described by an effective spin $S=1$ (which three components
describe states with zero, one and two bound-states)  the  topological order is similar to the spin one  Haldane  chain,  with  a spin $S=1/2$ localized at each edge. 
When trying to see whether our predictions are in agreement
 with these results, we face the problem  that they were  obtained at {\it half filling} where there is no spin-charge separation and therefore our approach does not strictly apply. If we assume though that  it does, thanks to the $SP(N)$ symmetry involved in these studies, the relevant bound-states are s-wave pairs belonging to the $ {\cal A}_{\mathbf{II}}$ class with $q=2$. 
At half-filling,  edge states in the charge sector are predicted  when the bound-state density  is an integer which, from (\ref{commcondbare}), implies $N/4$ to be integer. Even then,
the issue depends on the sign of the mass term $m$  in (\ref{diracBF}), so that for a given model both a trivial and a topological insulator may be stabilized depending on the mass parameter. This is exactly what has been first shown to happen  in Refs.(\cite{nonne1, nonne2}). Though we find this agreement  encouraging, it would be more satisfactory to check our predictions  to inquire whether these charge edge states exist in these systems for fillings  other than one-half. For instance, our analysis  opens the possibility of charge edge states for $N=6$, or spin $5/2$ fermions, at the filling $\bar \rho =1/3$. 
%
%

\section{Conclusions and open questions}
In this work we have  provided for a description of the low-energy
physics of interacting multi-species fermions in terms of the bound-states that are stabilized in these systems by the opening of a spin gap. We focused essentially on the massless charge degrees of freedom and on the  associated bound-state Luttinger liquid states. We have found that
a consistent bound-state Luttinger liquid state requires both
the charge $q$ and current $j$ quantum numbers, defining its  zero-mode spectrum,  to  satisfy the constraint $qj=N$ or
equivalently $qP_F= Nk_F$ for local bound-states. The latter condition may be viewed as some  form of the Luttinger theorem\cite{luttingertheorem, yamanaka}. Indeed assuming from the outset a gapless phase, the Luttinger liquid state, the equation $qP_F= Nk_F$ shows that there exists a massless mode at $2P_F= 2jk_F$ where $j=N/q$ for local bound-state solutions. Among the solutions of the latter equation a small finite subset  (i.e.  five types) of {\it generic} bound-states were characterized in terms of the elementary fermions. In terms of the charge and current quantum numbers
they are $(q, j)= (N,1)$ for both $SU(N)$ baryons and $ {\cal A}_{\mathbf{III}}$ class, $(q, j)= (2,N/2)$ for both $ {\cal A}_{\mathbf{I}}$ and $ {\cal A}_{\mathbf{II}}$ classes when $N$ is even and finally $(q, j)= (1,N)$ for the $ {\cal A}_{\mathbf{I}}$
class when $N$ is odd. We found that these are likely to be stabilized in systems which display an enlarged symmetry at low-energies and are associated with emergent duality symmetries in the spin sector.  Our results  are in agreement with previous findings for  three types of bound-states that were identified in the attractive $SU(N)$ Hubbard model with $N=3$ and $N=4$\cite{rapp, capponi, azaria2, dukelsky2008, ulbricht} as well as for $SP(N)$ models\cite{phle, capponi2, roux2008, phlezn}  relevant to describe general spin $F=(N-1)/2$ fermionic cold atoms. They are  fermionic or bosonic $SU(N)$ baryonic singlets and bound-states of them as well  as s-wave pairing states associated with $SP(N)$ symmetry. An important output of the previous studies is that these states are stable against small symmetry breaking fields. On top of these bound-states we also predict two new types of generic  bound-states with $O(N)$ symmetry:  p-wave bosonic pairs when $N$ is even and  composite fermions for odd $N$.   To our knowledge, the latter bound-states have not yet been  observed. 

Apart from   these generic bound-states we also predict the possible existence of other type of bound-states. 
When $N \ge 8$ for example, out of the two dual solutions $(q, j)= (8,1)$ and $(q, j)= (2,4)$ which are either $SU(8)$ baryons or s-wave pairing of the $ {\cal A}_{\mathbf{II}}$ class with $SP(8)$ symmetry,  there is another solution with $(q, j)= (4,2)$ which is self-dual. Upon increasing $N$ more solutions can be found (not necessarily self-dual) that are not generic bound-states. These states might be stabilzed in systems which {\it do not} exhibit
a dynamical enlarged symmetry at low-energies. One way to   think about   these bound-state solutions is to regard them as bound-states  made of generic bound-states themselves. For instance, in a system with $N=qj$ species, one may build up a $SU(q)$-singlet baryons made of either $SO(j)$-singlets charge one composite fermions or charge $2$ p-wave pairs.  The total symmetry in such a situation is  then $SU(q) \times SO(j)$. For odd $j$  the bound-state   has a charge $q$ and the unit of current is $j$. For even $j$ the total charge of the bound-state would be $2q$ and the unit of current $j/2$. Other possibilities involving other combinations of generic bound-states are of course possible. The point in the above construction is that it requires a hierarchy of scale. In the example we just gave, the gap in the $SO(j)$ sector should be much greater than the one in the $SU(q)$ one which is  consistent with the fact that the symmetry is not dynamically enlarged. More generally, one may anticipate that
the bound-states which are expected to be generic in the sense of
the DSE mechanism might be the building blocks for  more general bound-states. 

Another important result of the present  work concerns the relation between the possible existence of topological insulating phases with  the nature of the bound-states. In particular, the fact that it is the bound-state densiity $\rho_{\rm BS} = j\rho/q$
that controlls commensuration effects with the lattice. This opens
the possibility of non-degenerate Mott phases when $j > q$
or, equivalently, when the Fermi momentum $P_F = j k_F$
associated with these bound-states is enlarged. The fact that these phases display topological order is a highly non trivial problem. We though gave arguments that, in the particular case
of the generic bound-states, zero-energy edge states in the charge sector {\it may} be stabilized for either p-wave or s-wave bosonic bound-states associated with the classes $ {\cal A}_{\mathbf{I}}$ and $ {\cal A}_{\mathbf{II}}$ when $N$ is even. 
We finally stress that  zero energy edge states could also be stabilized in the spin sector. In particular, this leaves  open
the question of the topological nature of the Mott phase
associated with the composite fermions  at
integer bound-state densities\cite{azariaberg}.

\acknowledgments
We greatfully acknowledge illuminating discussions with E. Altman, E. Berg, F. Crepin, B. Doucot and P. Lecheminant. We also want to thank E. Boulat, S. Capponi,  G. Roux and A. M. Tsvelik for  collaborations  related to this work. The author is greatfull to A. Auerbach and the Physics departement of the Technion for their
kind hospitality during the past academic year while parts of this work were completed.

\appendix
\section{Bosonization Conventions}
In this appendix we discuss our bosonization conventions.
We recall the bosonized expressions of the elementary fermions
\be
c_{a,i}/\sqrt{a_0} = \frac{\kappa_a}{\sqrt{2 \pi }} [\re^{-i (k_F x + 2 \sqrt{\pi} \phi_{a, L})}+  \re^{i (k_F x + 2 \sqrt{\pi} \phi_{a, R})}], 
\label{bosoapp}
\ee
where the $\kappa_{a=1,...,N}$ anticommuting Majorana fermions,  $\{\kappa_a, \kappa_b\} = 2\delta_{ab}$, that insure the anticommutation between fermions of different specy. The
bosonic fields $\phi_{a, L}$ and $\phi_{a, R}$ do not commute and their commutators are given by $[\phi_{a, L}, \phi_{b, R}]= \mp i\delta_{ab}/4$ in order to insure the anticommutation between $L$ and $R$ fermions of the same species. The above commutators emerge in general  through the quantity
\be
 \gamma=\re^{-2\pi [\phi_{a,L}, \phi_{a,R}]},
\label{gamma}
\ee
which takes the values $\gamma = \pm i$. 
We find important to keep it explicit in the bosonization expressions in order to discuss parity issues as  under ${\cal P}$
\be
\phi_{a,L} \leftrightarrow - \phi_{a,R} ,  \gamma \rightarrow \gamma^*=-\gamma.
\ee
In this paper we make use of a basis in which spin and charge degrees of freedom  are described by  charge   bosonic fields $\Phi_c$ and $\Theta_c$ as well as  $N-1$ components spin bosonic fields $\vec \Phi$ and $\vec \Theta$ such that
\bea
\phi_a &=& \frac{1}{\sqrt{N}} \Phi_c + \vec \omega_a \cdot \vec \Phi \nonumber \\
\theta_a &=& \frac{1}{\sqrt{N}} \Theta_c + \vec \omega_a \cdot \vec \Theta 
\label{phiphispinapp}
\eea
where the $N$ vectors $\vec \omega_{a=1,...,N}$ are not independent and satisfy $\sum_{a=1}^N \vec \omega_a = 0$ together with  $\vec \omega_a \cdot  \vec \omega_b = \delta_{ab} -1/N$. 
With these definitions the parity symmetry act on the spin fields as
\be
{\cal P}: \vec \Phi \rightarrow - \vec \Phi, \; 
\vec \Theta \rightarrow  \vec \Theta, \;
 \gamma \rightarrow \gamma^*.
\label{parityspin}
\ee

The $SU(N)_1$ currents ${\cal I}_{L(R)}^A$ of Eqs.(\ref{suncurrents}) can be expressed in terms of the spin fields $\vec \Phi$ and $\vec \Theta$. Among them,  the $N-1$ Cartan generators of $SU(N)$ take a simple form. 
In each chirality sector  one has
\bea
\vec h_{L(R)} = \partial_x \vec \Phi_{L(R)}/\sqrt{\pi} = \sum_{a=1}^N  \vec \omega_a \; \partial_x  \phi_{a,L(R)}/\sqrt{\pi}
\eea
from which one deduces that 
\be
\vec h = \int dx\;  (\vec h_{L} +\vec h_{R} ) = \sum_{a=1}^N  \vec \omega_a Q_a
\label{cartanapp}
\ee
as given in (\ref{cartan}).

- {\it $SU(N)_1$ primaries operators.}

Following Affleck\cite{affleck} we define the $SU(N)_1$ primaries
operators $\Phi^{(m)}$ as
\be
\psi_{a_1, L}^{\dagger}...\psi_{a_{m}, L}^{\dagger}
\psi_{b_1, R}^{}...\psi_{b_m, R}^{} = \Phi^{(m)}_{a,b}\; \re^{im\sqrt{4\pi/N} \Phi_c}
\label{primarysun}
\ee
Using the bosonization formula (\ref{bosoapp}) as well as (\ref{phiphispinapp}) one may obtain the expression of 
$\Phi^{(m)}$ in terms of the spin fields $\vec \Phi$ and $\vec \Theta$. For instance,
the trace of $\Phi^{(m)}$ is given  by
\be
{\rm Tr}(\Phi^{(m)}) = \gamma^m \; \frac{\Gamma_m^2}{(2\pi)^m} \sum_{\{\vec \lambda_m\}} \re^{i\sqrt{4 \pi} \vec \lambda_m \cdot \vec \Phi}
\label{tracephim}
\ee
where $\vec \lambda_m = \sum_{j=1}^m \vec \omega_{a_j}$ and
$\Gamma_m = \Pi_{a=1}^m \kappa_a$. In 
(\ref{tracephim}) the sum runs over independent permutations
of the set $\{a_j\}$ compatible with the antisymmetry of (\ref{primarysun}). These operators  have the scaling dimension $d_m=\vec \lambda_m^2 =m(1-m/N)$ and transform under
parity as: ${\cal P}:  {\rm Tr}(\Phi^{(m)})  \rightarrow {\rm Tr}(\Phi^{(m)}) ^*$.

- {\it Duals of the $SU(N)_1$ primaries operators.} 
These objects appear natually in discussing both classes ${\cal A}_{\mathbf{I}}$ and $ {\cal A}_{\mathbf{III}}$ of dualities. In the case of the class ${\cal A}_{\mathbf{II}}$  we find it more
convenient to rely on the CFT embedding $SU(N)_1 \sim SP(N)_1 \times \mathbb{Z}_{N/2} $ which is discussed in details
in Ref.(\cite{boulat}) to which we refer. Both ${\cal A}_{\mathbf{I}}$ and $ {\cal A}_{\mathbf{III}}$ duality transformations have a simple
representation in terms of the spin fields $\vec \Phi$ and $\vec \Theta$. For ${\cal A}_{\mathbf{I}}$ we have
\bea
\vec \Phi \leftrightarrow \vec \Theta, \; 
\gamma \rightarrow \gamma^*,
\label{dualityvertexA1}
\eea
while for $ {\cal A}_{\mathbf{III}}$
\bea
\vec \Phi \rightarrow \vec \Phi  + \frac{\sqrt{\pi}}{2} \;  \vec e_p, \; 
\vec \Theta \rightarrow \vec \Theta - \frac{\sqrt{\pi}}{2} \;  \vec e_p,\; 
\gamma \rightarrow \gamma,
\label{dualityvertexA3}
\eea
where  $\vec e_p = \sum_{j=1}^p \vec \omega_j$. Using the above representations we find for the duals of ${\rm Tr}(\Phi^{(m)})$ 
\bea
{\cal A}_{\mathbf{I}}: \; {\rm Tr}(\widehat{\Phi^{(m)}}) &=& (\gamma^*)^m \;  \; \frac{\Gamma_m^2}{(2\pi)^m} \sum_{\{\vec \lambda_m\}} \re^{i\sqrt{4 \pi} \vec \lambda_m \cdot \vec \Theta},
\label{tracemA1} 
\\
{\cal A}_{\mathbf{III}}: \; {\rm Tr}(\widehat{\Phi^{(m)}})   &=& \gamma^m \; \frac{\Gamma_m^2}{(2\pi)^m} \sum_{\{\vec \lambda_m\}} \re^{i\sqrt{4 \pi} \vec \lambda_m \cdot \vec \Phi - i \pi \vec \lambda_m \cdot \vec e_p}.
\label{tracemA3}
\eea
In the particular case of the ${\cal A}_{\mathbf{I}}$ duality class   we find that under the parity transformation (\ref{parityspin})    
${\cal P}: {\rm Tr}(\widehat{\Phi^{(m)}}) \rightarrow (-1)^m {\rm Tr}(\widehat{\Phi^{(m)}})$ and that consequently the combination $\gamma^m {\rm Tr}(\widehat{\Phi^{(m)}})$ is ${\cal P}$ invariant. When  $m=N/2$ the latter quantity $\gamma^{N/2} {\rm Tr}(\widehat{\Phi^{(N/2)}})$ is also real. The reason is that  in (\ref{tracemA1})  the sum over $\{\vec \lambda_{N/2}\}$
contains both configurations  $\pm \vec \lambda_{N/2}$ thanks
to the property $\sum_{j=1}^N \vec \omega_j =0$.


\end{document}